\documentclass[11pt]{article}
\usepackage[a4paper, left=2cm, right=2cm, top=2.5cm, heightrounded]{geometry}
\usepackage[utf8]{inputenc}
\usepackage{multirow}
\usepackage{csvsimple}
\usepackage{graphicx}
\usepackage{verbatim}
\usepackage{enumerate}
\usepackage{amsmath}
\usepackage{caption}
\usepackage{amsfonts}
\usepackage{amsmath}{}
\usepackage{amsmath}{}

\usepackage{nicefrac}
\usepackage{float}
\usepackage{enumitem}
\usepackage[dvipsnames]{xcolor}
\usepackage{bbm}
\usepackage{mathtools}
\DeclareCaptionType{equ}[][]
\usepackage[titletoc,toc,title]{appendix} 
\usepackage[doublespacing]{setspace} 
\usepackage{caption, subcaption} 
\usepackage{sectsty}
\usepackage{xurl} 
\usepackage{tabularx, booktabs} 
\usepackage[authoryear]{natbib} 
\usepackage[hidelinks]{hyperref} 
\usepackage{cleveref} 
\usepackage{lscape} 

\defcitealias{koijen2019demand}{KY (2019)}

\title{\LARGE{Ponzi Funds}\thanks{We are grateful to Malcolm Baker, Alexios Beveratos, Lauren Cohen, Pierre Collin-Dufresne, Damien Van Eeckhaute, Robin Greenwood, Sam Hanson, Sebastian Hillenbrand, Erik Stafford, Jeremy Stein, Adi Sunderam, Bence Toth,  as well as the seminar participants at the Columbia Workshop in New Empirical Finance 2024 and the Cambridge University Algorithmic Trading Society Quant Conference 2024 for valuable comments.}}

\author{Philippe van der Beck\thanks{Harvard Business School}, Jean-Philippe Bouchaud\thanks{Capital Fund Management, Ecole polytechnique and France Académie des Sciences}, Dario Villamaina\thanks{Capital Fund Management}.}
\date{\today}

\begin{document}
\maketitle
\begin{abstract}
\singlespacing
\noindent 
Many active funds hold concentrated portfolios. Flow-driven trading in these securities causes price pressure, which pushes up the funds' existing positions resulting in realized returns. We decompose fund returns into a price pressure (self-inflated) and a fundamental component and show that when allocating capital across funds, investors are unable to identify whether realized returns are self-inflated or fundamental. Because investors chase self-inflated fund returns at a high frequency, even short-lived impact meaningfully affects fund flows at longer time scales. The combination of price impact and return chasing causes an endogenous feedback loop and a reallocation of wealth to early fund investors, which unravels once the price pressure reverts. We find that flows chasing self-inflated returns predict bubbles in ETFs and their subsequent crashes, and lead to a \textit{daily} wealth reallocation of \$500 Million from ETFs alone. We provide a simple regulatory reporting measure -- fund illiquidity -- which captures a fund's potential for self-inflated returns.

\end{abstract}
\hspace{5pt}
\newpage
\sectionfont{\fontsize{14}{15}\selectfont}
\subsectionfont{\fontsize{12}{15}\selectfont}
\subsubsectionfont{\fontsize{12}{15}\selectfont}

\section{Introduction} 
The collapse of Archegos Capital Management\footnote{\href{https://www.wsj.com/articles/inside-archegoss-epic-meltdown-11617323530}{Inside Archegos Epic Meltdown? (WSJ, 2021)}. Other episodes, such as the Gamestop saga, show how important flow-driven returns can be at the individual stock level (\cite{greenwood2023stock}).} prominently showed that when investment funds trade concentrated positions, portfolio returns can be driven by the funds' own price impact. Market participants seem unable to identify when realized returns come from price impact as opposed to fundamental determinants. Is this an isolated phenomenon or does investors' inability to differentiate self-inflated from fundamental returns have broader implications for the active investment management industry as a whole? 

A fast-growing literature investigates price pressure in financial markets and finds that non-fundamental demand shocks caused by for example flows, mandates, or index reconstitutions have a quantitatively meaningful impact on the cross-section of realized returns.\footnote{See e.g. \cite{shleifer1986demand}, \cite{harris1986price}, \cite{chen2004price}, \cite{frazzini2008dumb}, \cite{koijen2019demand}, \cite{pavlova2023benchmarking}, \cite{schmickler2020identifying}, \cite{greenwood2005short}, \cite{lou2012flow}, \cite{gabaix2021search}, \cite{jansen2021long}, \cite{wurgler2002does}, \cite{bretscher2022institutional}, \cite{greenwood2023stock}, \cite{coval2007asset}.} An equally large literature investigates the sensitivity of fund flows to past performance and finds that fund investors chase past return realizations.\footnote{See e.g. \cite{ippolito1992consumer}, \cite{dannhauser2019flow}, \cite{goldstein2017investor}, \cite{chevalier1997risk}, \cite{sirri1998costly}, \cite{lynch2003investors}, \cite{huang2007participation}, \cite{brown1996tournaments}, \cite{spiegel2013mutual}} In this paper, we empirically decompose realized fund returns at a high frequency into a price pressure and a fundamental component and show that investors are unable to disentangle managerial skill from price pressure, i.e. they equally chase realized returns from price impact and fundamental determinants. This has profound implications for the cross-section of fund flows, fund performance, and the underlying security returns more broadly. When funds trade based on flows, they exert price pressure on the securities in their portfolios as shown by \cite{lou2012flow}. Building on this mechanism, we show that the flow-performance relationship causes endogenous price spirals in the spirit of \cite{brunnermeier2009market} if future fund flows chase price pressure. The price pressure is a realized return to earlier investors in the affected securities resulting in what we label `self-inflated returns'. The inability of investors to differentiate fundamental and self-inflated returns leads to flows chasing past price impact, which causes further price pressure and an endogenous capital reallocation across investors: Via their own price impact, active funds effectively reallocate capital from late to early investors. 
We emphasize that the title of this paper does \textit{not} suggest that concentrated investment funds are \textit{literal} Ponzi schemes as defined by the SEC: ``\textit{A Ponzi scheme is an investment fraud that involves the payment of purported returns to existing investors from funds contributed by new investors.}''\footnote{See \url{https://www.sec.gov/spotlight/enf-actions-Ponzi.shtml}.} Instead, the term `Ponzi funds' merely conveys the notion of self-inflated returns. The reallocation of capital happens indirectly via observable market prices instead of direct capital transfers as in \textit{true} Ponzi schemes. The SEC also states that ``\textit{Ponzi schemes inevitably collapse, most often when it becomes difficult to recruit new investors or when a large number of investors ask for their funds to be returned.}'' This is closer to our proposed mechanism, as the wealth reallocation from self-inflated returns unravels once the price impact in the underlying securities reverts and investors stop misinterpreting self-inflated returns as managerial skill.

To assess whether investor flows chase their own price impact, we proceed in four steps. We start by laying out a simple analytical expression for self-inflated fund returns. Self-inflated returns are given by fund flows interacted with fund illiquidity, scaled by price impact. Fund illiquidity is given by the (portfolio-weighted) average of the fund's ownership in the underlying securities relative to their daily volume. Interacted with fund-flows, it measures how much of daily volume a fund on average buys when it reinvests flows in its existing positions. Fund illiquidity can be further decomposed into the product of portfolio concentration and fund size. We show that most funds have either concentrated portfolios (small specialized funds) or are large (diversified index trackers). When concentrated funds become large, their fund illiquidity spikes giving rise to self-inflated returns. 

Next, we estimate the price impact of flow-induced trades using ETFs' arbitrage-induced trading as a laboratory. ETFs are ideally suited because their portfolio holdings and flows are observable at a daily frequency, and because the vast majority of ETFs perfectly reinvest flows in their existing positions on the same day. A positive correlation between ETF flows and ETF returns can be due to i) price discovery or ii) price impact in the underlying securities from the flow-induced trades. Price discovery refers to the possibility that fund flows are not pure noise, but are contemporaneously correlated with fundamental news in the underlying. While raw fund flows may contain fundamental information about the ETFs' underlying holdings, self-inflated fund returns are driven by the interaction of flows with the illiquidity of the underlying. This decomposition allows for a difference-in-difference estimator, by conditioning the price impact of fund flows on fund illiquidity. Intuitively, if the correlation between returns and flows is stronger when fund illiquidity is high, this suggests price impact. We find that self-inflated returns explain 8\% of time-series in variation in returns for funds that are both large and hold concentrated portfolios. The importance of self-inflated returns in explaining funds' overall returns increases monotonically in both the size and the concentration of their portfolios.

Next, we investigate the performance sensitivity of fund flows at a daily frequency. We document a strong relationship between fund flows and fund returns at a daily frequency and find that the weight investors place on more distant fund returns decays exponentially. We then decompose past fund returns into a self-inflated return and a fundamental component. Quite strikingly, the flow-performance relationship is not confined to the fundamental component of fund returns. Instead, investors also chase the price impact component of fund returns. The coefficients on both return components are almost identical and both highly statistically significant. We conclude that investors are unable to differentiate between self-inflated returns (price impact) and fundamental returns (stock-picking skill). This has important implications for the wealth distribution of managed funds, to the extent that ETFs are representative of the broader investment management industry. Our findings place a relatively strong constraint on the rationality of ETF investors. While pure return chasing may indicate learning about the fund manager's skill (\cite{berk2004mutual}), we show that investors chase their own impact and are hence unable to distinguish between realized and expected returns.

Last, we combine the self-inflated returns with return chasing to quantify the economic importance endogenous price spirals caused by impact chasing: Funds with concentrated portfolios in illiquid basket securities have a high potential impact on the underlying. Flows into these funds pushes up the price of the underlying, leading to high realized fund returns. Following this price impact return, investors allocate more flows to these funds, which we label `Ponzi flows'. The economic magnitude of Ponzi flows and their price impact are meaningful. Around 2\% of all daily flows and 8-12\% of flows in the top decile of illiquid funds can be attributed to Ponzi flows. We estimate that every day around \$500 Million of investor wealth is reallocated because of the price impact of Ponzi flows. We furthermore find that funds with high Ponzi flows experience subsequent drawdowns of over 200\%. \\
\noindent \textbf{Related Literature.}
The performance of actively managed investment funds has been a widely studied area in financial economics.\footnote{See \cite{jensen1968performance}, \cite{carhart1997persistence}, \cite{pastor2002mutual} and \cite{cohen2005judging}} This is partly owed to the fact that flows and the performance of active fund managers give insights into the efficiency of financial markets, as well as the reaction of consumers to observable measures of product quality. Starting with \cite{ippolito1992consumer}, \cite{chevalier1997risk}, and \cite{sirri1998costly}, a large body of work examines the sensitivity of fund flows to past performance.\footnote{See \cite{huang2007participation}, \cite{goldstein2017investor}), \cite{jiang2018active}, \cite{berk2016assessing}, \cite{barber2016factors}, \cite{jin2022swing},  \cite{falato2021financial}.} If investors can rationally infer managers' skills from the history of realized returns, then capital efficiently flows to positive net present value opportunities (which are held by the most skilled managers). However, it is difficult to measure the skills of active money managers from past returns -- particularly over short horizons. \cite{berk2004mutual} provide a rational explanation for the lack of performance persistence in managed funds, which is driven by the fact that successful funds grow in size until their own price impact eliminates their profitable investment opportunities. We show that, when interpreting past return realizations, investors are unable to differentiate between manager skill and price impact. The constraint on the rationality of fund investors, paired with a large price impact of active funds, leads to Ponzi-like reallocations of capital among fund investors that unravels when the price impact in the underlying securities reverts.

Second, our paper is related to growing literature on linking institutional demand shocks and asset prices via estimating demand systems featuring downward-sloping demand curves (see \cite{koijen2019demand}, \cite{bretscher2022institutional}, \cite{koijen2020investors}, \cite{haddad2021competitive} and \cite{gabaix2021search}). In a related paper \cite{darmouni2022nonbank} study the interaction of return chasing and price impact during asset fire sales. The majority of papers in this stream of literature are studying the impact of demand shocks at a lower frequency. We contribute to these findings by estimating the impact of institutional demand shocks at a high (daily) frequency and find similar overall magnitudes. The high-frequency nature of our demand shocks allows us to investigate the permanent versus transitory nature of price impact and how it aggregates over time. We provide a direct link to price impact estimates at a higher frequency (see e.g. \cite{chacko2008price}, \cite{toth2011anomalous}, \cite{frazzini2018trading}, \cite{bouchaud2018trades}, \cite{kyle2016market}) and show that a square root specification and scaling demand shocks by daily volatility strongly dominates the linear specification at a daily frequency. We find that around 50\% of the initial daily price impact from flow-induced trades reverts in the subsequent 5-10 days. This is very much in line with the results of \cite{bucci2018slow} for the relaxation of impact on single stocks, using a completely different data set.

Last, our paper is also related to the growing literature on ETFs (see \cite{ben2018etfs}, \cite{glosten2021etf}, \cite{ben2023competition}, \cite{box2021intraday}, \cite{dannhauser2019flow}). On the one hand, ETFs may improve price discovery in the underlying basket securities by offering investors superior liquidity.\footnote{See \href{https://www.blackrock.com/au/intermediaries/ishares/etf-market-realities}{Blackrock ETF Market Realities}} On the other hand, the liquidity of ETFs may attract a new clientele which introduces non-fundamental volatility in the underlying basket securities (\cite{ben2018etfs}). In this paper, we disentangle the two channels and argue that the nature of ETFs -- a liquid vehicle that tracks a potentially illiquid basket of securities -- is causing considerable price distortions in the underlying securities. Specifically, we argue that when ETFs hold concentrated positions in individual stocks, they accumulate substantial ownership in these securities as the ETF grows in size. The ETF arbitrage mechanism propagates ETF demand in the underlying securities causing a non-fundamental price impact, which enhances the ETF's own return. ETF investors are return chasing and cannot differentiate the mechanical price impact due to ETF arbitrage from fund manager skill. They hence purchase further ETF shares leading to continued price impact, further ETF share issuance, and endogenous boom and bust cycles in both the ETF price and the underlying. 

The remainder of this paper is structured as follows. Sections \ref{Case Study} and \ref{Data} provides a case study of an anonymized thematic ETF that motivates our study and a brief description of our main data sources. The remaining sections closely follow the four steps outlined above. Section \ref{Model} provides an analytical expression for self-inflated fund returns and introduces the concepts of fund illiquidity and portfolio concentration. In section \ref{Estimation} we estimate self-inflated returns. Section \ref{Impact Chasing Section} tests whether investors chase past self-inflated returns. Section \ref{Self-Inflated Feedback Loops} estimates the economic importance of the self-inflated feedback loop and provides applications. Section \ref{Conclusion} concludes.

\section{Case Study: The price impact of a large thematic ETF}
\label{Case Study}
\begin{figure}[H]
\caption{\textbf{Flow-Driven Price Impact of a Large Thematic ETF.} The figure plots the cumulative return and cumulative price impact of an anonymized thematic ETF. Price impact (red line) is computed as the portfolio-weighted average of stock-specific price impact, summed cumulatively over time. Stock-specific price impact is modeled as $\sigma \sqrt{Q/V}$ where $\sigma$ and $V$ are daily volatility and volume, and $Q$ are the daily flow-driven trades of the ETF (see Section \ref{Model} for details on the modeling of price impact). The ETFs' cumulative return is the daily net asset value scaled by its January 2019 value.}
\centering
\includegraphics[scale=0.9]{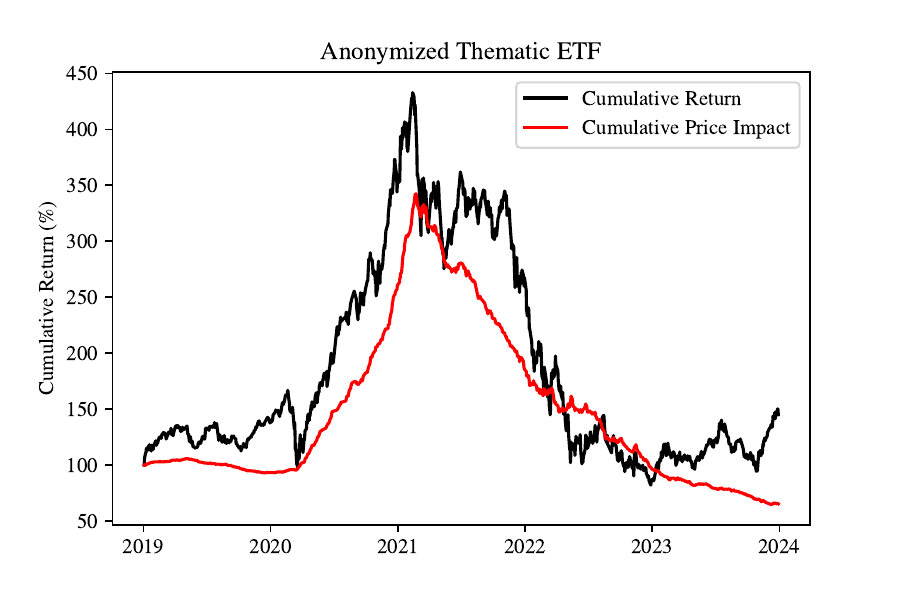}
\label{ETF collapse Example}
\end{figure}

For illustration, consider the simple case study of an anonymized thematic ETF, whose price quadrupled between March and August of 2021 and collapsed shortly after. Figure \ref{ETF collapse Example} plots the cumulative daily price impact of the thematic ETF's flow-induced trades against its cumulative return.\footnote{The price impact estimation is explained in the following sections. Importantly, it does not use prices or returns and is only calculated using daily share volumes, past volatility, and flow-driven purchases. Our results remain unchanged if we, instead of estimating our own price impact, take existing price impact calibrations/estimations from both the micro-structure (see e.g. \cite{hasbrouck2007empirical}) and asset pricing literature (see e.g. \cite{shleifer1986demand}).}

The figure shows a strikingly high \textit{quantitative} resemblance of the thematic ETF's cumulative price impact and its return. Its raw daily returns and flow-induced trades are over 40\% correlated. During this period, the thematic ETF's positions were 20 times larger than the daily dollar volume in those securities. This implies that whenever the thematic ETF received a $1\%$ inflow on a certain day and proportionally rescaled its positions, it bought $20\%$ of the daily volume in the underlying stocks. Because its portfolio was heavily concentrated in these individual securities and it received over 200\% inflows, a large portion of its portfolio return was driven by its own price impact.

\section{Data}
\label{Data}
 We obtain daily holdings of all ETFs in the US from 2019 to 2024 from ETF Global. The holdings data include the daily shareholdings of each ETF provider in the underlying securities, as well as the daily shares outstanding. This allows us to construct a host of daily portfolio-level variables, such as flows, liquidity of the underlying portfolio, and flow-driven trades. Security prices and characteristics are from CRSP and Compustat. In order to minimize potential data errors, we restrict the universe of funds to US 1500 most liquid equity funds, computed point-in-time. Neverthless we have conducted multiple explorations, increasing the pool of funds both geographically and in terms of liquidity, and observed only minor changes.

 Table \ref{Summary Statistics} provides summary statistics of the ETF data. Our main analysis focuses on ETF data, for which daily holdings are available. While ETFs are exchange-traded vehicles that generally attempt to replicate a rules-based index, the portfolios of \textit{thematic} ETFs are subjectively chosen by a fund manager. Our sample consists of 1868 ETFs with 1,175,487 ETF-day observations and 34,897,091 ETF-day-stock observations. The median ETF has \$150 million dollars under management, holds 62 stocks in its portfolio, has an active share of 31\% (i.e. it tilts 31\% of its assets away from the value-weighted market portfolio), receives a daily inflow of \$1.6 Million (0.13\% relative to assets). 

\begin{table}[h]
\center
\caption{\textbf{Summary Statistics}.\\ The table reports summary statistics of the sample of daily ETF holdings from 2019 to 2024. Active share is computed as relative to the value-weighted portfolio of all stocks in the ETFs universe and Industry HHI measures the industry tilt of ETFs, i.e. $\sum_j (w_{i,t}(j)-w_t^m(j))^2$, where $w_{i,t}(j)$ and $w_t^m(j)$ are the weights of the ETF and the market portfolio in industry $j$ respectively. Portfolio illiquidity $\mathcal{I}$, concentration $\mathcal{C}$ and size $\mathcal{S}$ are defined in Section \ref{Model}). Daily alphas are computed with respect to the CRSP total stock market return.}
\begin{tabular}{lrrrrrrr}
\toprule
{} &    Mean &     Std &      Min &     Q1 &  Median &      Q3 &      Max \\
\midrule
AUM (\$ Billion)        &    2.45 &   14.99 &     0.00 &   0.03 &    0.15 &    0.81 &   498.40 \\
Daily Flow (\% AUM)     &    0.44 &    5.02 &   -36.85 &  -0.45 &    0.13 &    0.83 &    47.69 \\
Daily Flow (\$ Million) &    4.17 &  175.12 & -4948.75 &  -4.47 &    1.60 &    8.29 &  4963.34 \\
Daily Return (\%)       &    0.01 &    2.12 &  -767.68 &  -0.68 &    0.06 &    0.77 &    57.13 \\
Daily Alpha (\%)        &   -0.05 &    1.03 &   -99.74 &  -0.41 &   -0.04 &    0.31 &    54.04 \\
Number of Stocks       &  142.68 &  200.32 &     1.00 &  29.00 &   62.00 &  155.00 &  1538.00 \\
Active Share           &    0.31 &    0.20 &     0.00 &   0.13 &    0.31 &    0.47 &     0.94 \\
Industry HHI           &    0.24 &    0.30 &     0.00 &   0.03 &    0.09 &    0.34 &     0.97 \\
Fund Illiquidity $ \mathcal{I}$          &    0.19 &    1.54 &     0.00 &   0.00 &    0.02 &    0.09 &   325.76 \\
Portfolio Concentration $\mathcal{C}$          &    1.75 &    2.08 &     0.00 &   0.35 &    0.82 &    1.98 &   418.58 \\
Fund Size $\mathcal{S}$          &    0.36 &    4.40 &     0.00 &   0.00 &    0.02 &    0.12 &   453.49 \\
\bottomrule
\end{tabular}

\label{Summary Statistics}
\end{table}

\section{Price Impact and Self-Inflated Fund Returns}
\label{Model}
In this section, we define fund-level illiquidity and show that self-inflated fund returns can mathematically be expressed as the product of fund flows and illiquidity.  

\subsection*{Notation} The setup closely follows \cite{pastor2020fund}. There are $n=1,...,N$ underlying securities with price $P_{n,t}$ at time $t$, realized return $r_{n,t+1}$, and a fund $i$ with assets $A_{i,t}$ that tracks a portfolio of securities. The fund's portfolio weight in stock $n$ is $w_{i,n,t}$ and the total portfolio return (in the absence of management fees) is $R_{i,t+1} = \sum_n w_{i,n,t} r_{n,t+1}$. Dollar flows in the fund between $t$ and $t+1$ are denoted by $F_{i,t+1}$, and flows relative to assets are given by $f_{i,t+1}=F_{i,t+1}/A_{i,t}$. We define the flow-driven trade $Q_{i,n,t}=w_{i,n,t-1}F_{i,t}$ as the dollar trade of fund $i$ in stock $n$ at time $t$ resulting from reinvesting flows in proportion to existing portfolio weights.

\subsection*{Price Impact}
For simplicity of exposition, we omit the fund-level indicator $i$ when there is no ambiguity. Flow-driven trading $Q_{n,t}$ in the underlying securities causes a price impact. We assume that the price impact is larger when trading a larger fraction of $n$'s average daily dollar volume $V_{n,t}$. The main results in the paper remain unchanged when we scale demand shocks by other measures of liquidity such as float-adjusted market cap. Following the literature on transaction costs (see e.g. \cite{chacko2008price}, \cite{frazzini2018trading}, \cite{toth2011anomalous}, \cite{kyle2016market}) we model price impact as 
\begin{equation}
\label{Price Impact}
    \text{Price Impact}_{n,t} = \theta \sigma_{n}\bigg(\frac{Q_{n,t}}{V_{n,t}}\bigg)^{\eta}
\end{equation}
The price impact (expressed as a return) is proportional to the volatility of the stock $\sigma_{n}$ and the trade relative to supply $\big(\frac{Q_{n,t}}{V_{n,t}}\big)^{\eta}$. We allow for non-linearity $\eta$ in the relationship between trade size and price impact. A large empirical literature estimates the price impact of trades at a high frequency at finds a square root impact $\eta \approx 0.5$ and a $\theta$ of order unity.\footnote{See \cite{bouchaud2018trades} for a detailed literature summary.} The literature on flow-driven trading at a quarterly frequency (see e.g. \cite{lou2012flow}) typically assumes a linear specification $\eta=1$, scaling demand shocks by market cap, and omitting the volatility pre-factor.\footnote{Formally, the linear price impact specification in e.g. \cite{lou2012flow}, \cite{pavlova2023benchmarking}, \cite{koijen2019demand} is given by $\text{Price Impact}_{n,t} = \theta \frac{Q_{i,n,t}}{M_{n,t}}$ where $M_{n,t}$ is the market capitalization of $n$.} Defining price impact in this way leaves the \textit{qualitative} results of the paper unchanged. Figures \ref{HorseRace Fund-Level}, \ref{HorseRace Stock-Level}, and \ref{FIT Horserace} in the Appendix show that (\ref{Price Impact}) strongly dominates the linear model in a horse race at different levels of aggregation.

\subsection*{Fund Illiquidity, Portfolio Concentration, and Size}
\begin{figure}[H]
\centering
\caption{\textbf{Portfolio Concentration versus Fund Size in ETFs}. The figure plots the average portfolio concentration $\mathcal{C}_{t}$ and size $\mathcal{S}_{t}$ for each ETF in our sample. Each dot represents a single ETF. Portfolio concentration (on the x-axis) is $\mathcal{C}_{t}=\sum_n w_{n,t}^{1+\eta}/v_{n,t}^\eta$ where $v_{n,t}=\tilde{V}_{n,t}/\sum_n \tilde{V}_{n,t}$ are hypothetical liquidity weights and $\tilde{V}_{n,t} = V_{n,t}\sigma_n^{-1/\eta}$ is effective liquidity. Fund size (on the y-axis) $\mathcal{S}_{t}=A_t/\sum_n \tilde{V}_{n,t}$ is total assets relative to the total liquidity of the underlying securities. We choose $\eta=0.5$ for the computation of all measures. The labeled dots represent illustrative examples.}
\center
    \includegraphics[scale=0.9]{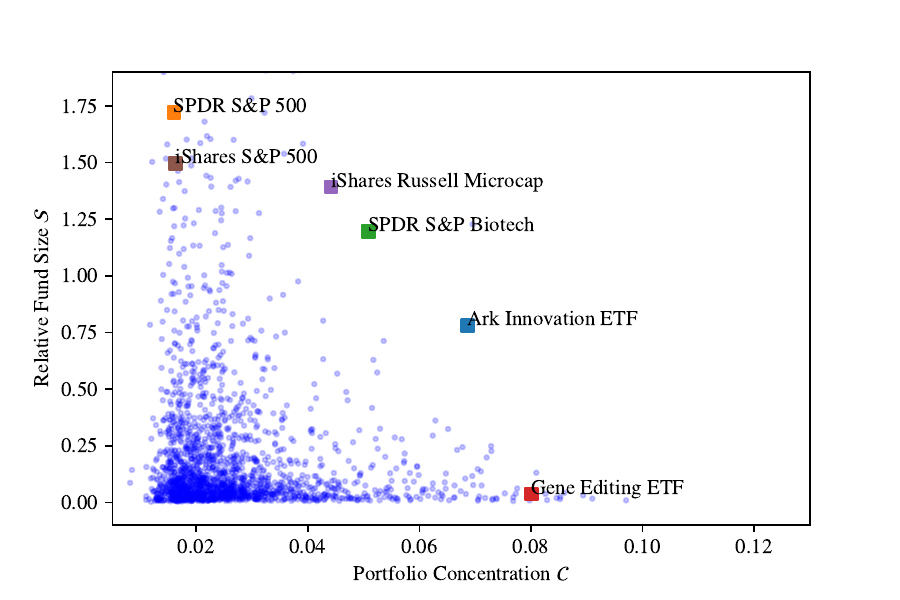}
    \label{Concentration versus Size}
\end{figure}

The dollar position of a fund in $n$ is given by $w_{n,t}A_t$. We define the position-level illiquidity as
\begin{equation}
    \mathcal{I}_{n,t} \equiv \sigma_n\bigg(\frac{w_{n,t}A_t}{V_{n,t}}\bigg)^\eta.
\end{equation}
 It is the dollar position relative to underlying liquidity. For example, as of Tuesday April 23rd 2024, the ARK Technology ETF held a \$525 Million position in Roku Inc., which has an average daily volume of \$150 Million and an average daily volatility of 1.3\%, resulting in $\mathcal{I}_{n,t}\approx 0.05$.\footnote{Using $\eta=1$, we have that $\mathcal{I}=0.013\times\frac{525}{150}$. See \url{https://ark-funds.com/funds/arkk/}}. We can decompose position-level illiquidity in a component that is related to the portfolio weight (i.e. the concentratedness of the fund's position) and the assets under management (i.e. size of the fund). Position-level illiquidity can be further decomposed into the product of position-level concentration $\mathcal{C}_{n,t}$ and fund size $\mathcal{S}_{t}$
 \begin{equation}
     \mathcal{I}_{n,t} = \mathcal{C}_{n,t} \times \mathcal{S}_{t}
 \end{equation}
In the spirit of \cite{pastor2020fund} we define concentration $\mathcal{C}_{n,t}$ as a measure of how strongly the portfolio weight deviates from a liquid portfolio. Formally, $\mathcal{C}_{n,t} \equiv \big(\frac{w_{n,t}}{v_{n,t}}\big)^\eta$ where $v_{n,t}=\frac{\tilde{V}_{n,t}}{\sum_n \tilde{V}_{n,t}}$ are hypothetical liquidity weights within the fund's universe and $\tilde{V}_{n,t} = V_{n,t}\sigma_n^{-1/\eta}$ is effective liquidity.
Interestingly, when $\eta=1$, effective liquidity $\frac{V_{n,t}}{\sigma_n}$ is of the same order of magnitude as market capitalization.\footnote{Daily volatility ($\sim 2 \%$) and daily volume ($\sim 0.5 \%$ of market cap) are of the same order of magnitude. As noted in \cite{bouchaud2022inelastic}, this actually provides an alternative explanation for the fact that the multiplier in \cite{gabaix2021search} is $O(1)$.} Therefore it is not too surprising that the estimated $\theta$ from low and high-frequency studies tend to be of similar magnitude despite the different scaling (see Section \ref{Estimation} for details.) Fund size $\mathcal{S}_{t}\equiv \frac{A_t}{\sum_n \tilde{V}_{n,t}}$ are the assets under management relative to the total effective liquidity of the underlying portfolio. The position-level variables $\mathcal{I}_{n,t}$ and $\mathcal{C}_{n,t}$ can be summed across the portfolio to arrive at fund-level illiquidity $\mathcal{I}_{t}\equiv\sum_n w_{n,t}\mathcal{I}_{n,t}$ and portfolio-concentration $\mathcal{C}_{t}\equiv\sum_n w_{n,t}\mathcal{C}_{n,t}$. The decomposition of illiquidity into concentration and size also holds at the fund-level
 \begin{equation}
     \mathcal{I}_{t} = \mathcal{C}_{t} \times \mathcal{S}_{t}
 \end{equation}
 Fund illiquidity is the average ownership of the fund relative to the liquidity of its underlying portfolio constituents.\footnote{Formally $\mathcal{I}_{t} = \bigg(\sum_n \sigma_{n,t}w_{n,t}\big(\frac{w_{n,t}A_t}{V_{n,t}}\big)^{\eta}\bigg)^{1/\eta}$} In the linear low-frequency specification (i.e. when $\eta=1$, supply is equal to market equity, and the prefactor is one) $\mathcal{C}_{t}=\sum_{n} \frac{w_{n,t}^2}{m_{n,t}}$ where $m_{n,t}$ are market weights as in \cite{pastor2020fund}. In this case $\mathcal{I}_{t}=\sum_n w_{n,t}z_{n,t}$ where $z_{n,t}$ is the fraction of shares outstanding held by the fund. $\mathcal{I}_{t}$ is higher for portfolios with a high weight in assets with a small market cap. Thus $\mathcal{I}_{t}$ is conceptually similar to the active share by \cite{cremers2009active}, but takes into account the supply of the underlying by dividing by (as opposed to subtracting) market weights. Figure \ref{Concentration versus Size} plots the average portfolio concentration $\mathcal{C}_{t}$ and size $\mathcal{S}_{t}$ for each ETF in our sample.
In line with the findings in \cite{pastor2020fund}, larger ETFs (such as the SPDR S\&P500 ETF) typically hold less concentrated portfolios and ETFs with the most concentrated portfolio (such as the Kelly Gene Editing ETF) are small relative to the volume of the stocks they hold. The ETFs that are located close to the axes have a low fund illiquidity $\mathcal{I}_t$ and are therefore not affected by self-inflated fund returns. All ETFs that are located towards the center of the figure have a high fund illiquidity and are prone to self-inflated fund returns. The next section describes the relationship between fund illiquidity and self-inflated returns. 

\subsection*{Self-Inflated Fund Returns}
Price impact from flow-driven trading is a realized return on the existing positions of the fund. The realized fund return due to flow-driven price impact is given by the portfolio weighted sum of stock-specific price impact $\sum_n w_{n,t-1} \text{Price Impact}_{n,t}$. Plugging in (\ref{Price Impact}) and $Q_{n,t}=w_{n,t-1}F_{t}$ yields the self-inflated fund return $R^{\mathcal{I}}_{t}$:
\begin{equation}
\label{Self-Inflated Return Equation}
    R^{\mathcal{I}}_{t} =  \theta f_t^{\eta}\mathcal{I}_{t-1}
\end{equation}
where $\mathcal{I}_{t}$ is the \textit{fund-level illiquidity} defined above. The self-inflated fund return $R^{\mathcal{I}}_{t}$ is given by the relative flow in the fund $f_{t}^{\eta}$, interacted with fund illiquidity $\mathcal{I}_{t-1}$, scaled by price impact $\theta$.\footnote{For simplicity of exposition, we assume that flows $f_{t}$ are positive. In the presence of outflows ($f_{t}<0$), which is the empirically relevant case, the expression becomes $ R^{\mathcal{I}}_{t} =  \theta \text{sign}(f_{t})|f|^\eta_{t}\mathcal{I}^{\eta}_{t-1}$.} In essence, $\mathcal{I}_{t}$ measures the \textit{potential} impact of a fund on the underlying portfolio constituents: The more concentrated the fund's portfolio weights are relative to the underlying supply, the greater the potential impact of a 1\%  flow in the fund. Naturally, if a fund holds a concentrated portfolio, not only are the trades in individual securities larger, but the price impact on individual securities also has a greater effect on the entire fund return. As shown above, fund illiquidity is driven by the size of the fund relative to its underlying constituents, as well as the portfolio tilts within the fund's universe. \cite{pastor2020fund} find, that larger funds typically hold less concentrated portfolios which results in moderate values of $\mathcal{I}_{t}$ in the cross-section of managed funds.\footnote{Empirically, the largest funds are index funds which hold market weights and therefore tilt towards the most liquid securities within their portfolio. Therefore their high $\mathcal{S}_{t}$ is offset by a low $\mathcal{C}_{t}$.} However, when a concentrated fund attracts significant inflows and becomes large without readjusting its portfolio towards more liquid securities, it will have a high $\mathcal{I}_{t}$ and be subject to self-inflated fund returns.

\section{Estimating Self-Inflated Returns}
\label{Estimation}
The difficulty in estimating price impact lies in the fact that trades likely contain information about the underlying fundamentals or risk exposures. They are correlated with unobserved variation in prices (the error term) leading to biased price impact estimates. At a lower frequency, the empirical literature on low-frequency price impact has therefore carefully constructed exogenous variation in demand from e.g. index inclusions (\cite{shleifer1986demand}), mutual fund flows (\cite{lou2012flow}), and dividend reinvestments (\citet{hartzmark2022predictable}). Building on this literature, we estimate high-frequency price impact from flow-driven trades by ETFs. While the exchange-traded nature of ETFs slightly alters the construction of flows and flow-driven trading (see Appendix \label{ETFs versus Mutual Funds} for details), the underlying mechanics of flow-driven trading in mutual funds and ETFs are identical. The advantage of using ETF data is that the vast majority of ETFs perfectly reinvest flows in their existing positions on the same day. The timing of the reinvestment is more difficult to confirm for mutual funds as portfolio holdings are reported at a monthly or quarterly frequency.

\subsection{Qualitative Evidence from Portfolio Sorts}
Before diving into the quantitative estimation, we first provide some intuitive evidence for the existence of self-inflated fund returns from simple portfolio sorts. Every day, we sort all 1868 ETFs into decile buckets based on their inflows. Within each bucket, we split the sample into ETFs with a high fund illiquidity ($\mathcal{I}_{t}$ in top decile) and all other funds. We then compute the average return across ETFs for each bucket. 
\begin{figure}[H]
\centering
\caption{\textbf{Daily ETF Returns and Inflows}. The figure plots the daily average return of ETFs across different flow buckets. Every day, we sort ETFs into deciles based on their inflows and compute their average return. Within each bucket, we split the sample into ETFs with a high fund illiquidity, for which $\mathcal{I}_{t}$ is in the top decile (red bars), and all other funds (green bars).}
\center
    \includegraphics[scale=0.9]{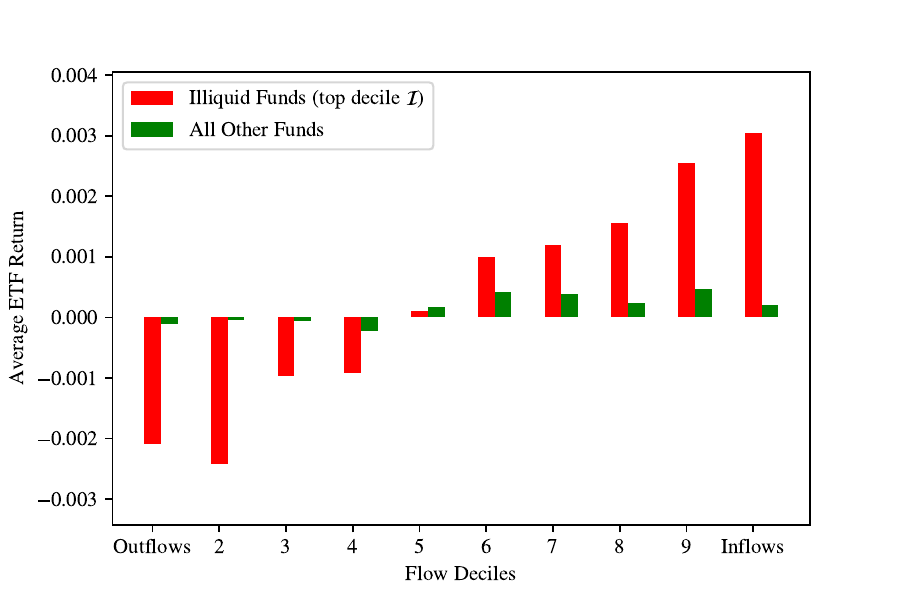}
    \label{Portfolio Sorts}
\end{figure}
When the funds hold a liquid portfolio, there is virtually no relationship between daily flows and returns. However, for the top illiquid funds, the realized return increases monotonically with the flow. In Appendix \ref{Portfolio Sorts Robustness} we show that this pattern holds for raw returns, excess returns, and abnormal returns, as well as for different sorts and definitions of flows and illiquidity. While this is preliminary \textit{qualitative} evidence for the hypothesis that fund flows affect fund-level returns if the funds hold illiquid portfolios, it does not get at the \textit{quantitative} magnitude. The next section estimates the quantitative link ($\theta$), using ordinary least squares in a difference-in-difference setting. 

\subsection{Price Impact at the Stock Level}
For simplicity of exposition we focus on the estimation of $\theta$ and take and take the square root functional form ($\eta=0.5$) as given.\footnote{In Appendix Figure \ref{HorseRace Fund-Level} we show that the square root specification dominates the linear specification in terms of statistical significance and incremental $R^2$. We also estimate $\eta \approx 0.6$ in a joint nonlinear panel estimation of $\theta$ and $\eta$. The results are available upon request.} 
On a given day $t$, the return $r_{n,t}$ on stock $n$ is driven by a flow-driven trading component, and a residual component $r^{\perp}_{n,t}$ capturing for example fundamental news or risk factor exposures. Formally, $r_{n,t} = \theta f^\eta_{t} \mathcal{I}_{n,t-1} + r^{\perp}_{n,t}$. A potential concern with using flow-driven trading as exogenous variation in investor demand is that the flows themselves may contain information about the underlying securities held by the fund. This may be particularly worrisome for ETFs, which are precisely used by investors to take theme and sector-specific bets. We, therefore, propose to estimate price impact in a pooled regression across fund-stock-time observations as opposed to aggregating flow-driven trades across funds. This allows us to control for the information in flows using fund-time fixed effects $\alpha_{i,t}$:
\begin{equation}
\label{Estimation Equation}
    r_{n,t} = \alpha_{i,t} + \beta \mathcal{I}_{i,n,t-1} + \theta f^\eta_{i,t} \mathcal{I}_{i,n,t-1} + \text{Controls} + \epsilon_{i,n,t}.
\end{equation}
The different funds in our sample are denoted by $i=1,...,I$. Note that portfolio tilts may contain information about the underlying resulting in a positive correlation of $\mathcal{I}_{i,n,t-1}$ and return determinants in the cross-section. The impact of flow-driven trading, however, comes from the interaction of position-level illiquidity and fund flows. Thus controlling for position-level illiquidity $\mathcal{I}_{i,n,t-1}$ results in an implicit difference-in-difference estimator. Intuitively, a positive correlation between flow-driven trades and returns could be i) because fund flows contain information, which is captured by $\alpha_{i,t}$, ii) because fund tilts contain information, which is captured by $\beta$, or iii) because of price impact, which is captured by $\theta$. We estimate \ref{Estimation Equation} over the panel of pooled ETF-stock-day observations from 2018 to 2024 using ordinary least squares. Table \ref{OLS Regression Results} reports the estimated coefficients. 
\begin{table}[H]
\center
\caption{\textbf{Price Impact at the Stock Level}. The table reports the estimated price impact coefficients from pooled OLS regressions using daily data on stock returns, ETF flows $f^{\eta}_{i,t}$, and position-level illiquidity $\mathcal{I}_{i,t-1,n}$. Specification (1) regresses daily stock returns onto flows. Specification (2) estimates the price impact using the interaction of illiquidity and flows $f^{\eta}_{i,t}\mathcal{I}_{i,t-1,n}$ including fund and time fixed effects. Specification (3) removes the fund-time fixed effects. Specification (4) adds concentration $\mathcal{C}_{i,t-1,n}$ and the interaction of concentration and flows $f^{\eta}_{i,t}\mathcal{C}_{i,t-1,n}$. We have fixed $\eta=0.5$ such that $f^{\eta}_{i,t}$ are square-root flows. For simplicity of exposition the table reports $f^{\eta}_{i,t}$ and $\mathcal{I}_{i,t-1,n}$ as $f$ and $\mathcal{I}$ respectively. Standard errors are double clustered at the day and the stock-day level.
}
\begin{tabular}{lllll}
\toprule
{} &      (1)      &      (2)      &        (3)       &        (4)       \\
                      &               &               &                  &                  \\
\midrule
const                 &  9.528e-07    &  -0.0002***   &  -0.0002***      &  -1.29e-05       \\
                      &  (0.0019)     &  (-31.531)    &  (-31.966)       &  (-0.0261)       \\
$f$               &  0.0013**     &  -8.424e-06   &                  &                  \\
                      &  (2.5481)     &  (-0.9349)    &                  &                  \\
$ \mathcal{I}$               &               &  -0.0244      &  -0.0308         &  0.0111          \\
                      &               &  (-1.2042)    &  (-1.0676)       &  (0.4932)        \\
$f\times \mathcal{I}$              &               &  0.3354***    &  0.3337***       &  0.6951***       \\
                      &               &  (7.5979)     &  (7.4173)        &  (7.3457)        \\
\midrule

Effects                 &                 &     Investor-Time    &  Investor-Time  &    \\
Triple Diff.            &                 &                 &           Yes      &  Yes         \\

No. Observations      &  34897091     &  34897091     &  34897091        &  34897091        \\
R-squared             &  7.341e-05    &  8.005e-05    &  8.927e-05       &  0.0008          \\
\bottomrule
\end{tabular}

\label{OLS Regression Results}
\end{table}
To set the stage, we initially estimate an ordinary regression of daily stock returns $r_{n,t}$ onto ETF flows $f_{i,t}$. The coefficient is close to 0, but statistically significant suggesting that there is some information contained in the fund-level flows. Next, we estimate the difference-in-difference specification via the interaction term with fund illiquidity $f_{i,t+1}^{\eta}\mathcal{I}_{i,n,t}$, controlling for fund-day fixed effects. The coefficient on the interaction term is around 0.33 and highly statistically significant (t-stat $>$10). It has the following structural interpretation: If a fund's average weighted ownership relative to supply is 50\% ($\mathcal{I}_{i,n,t}=0.5$), a 10\% inflow causes an additional ETF return of $0.33 \times 10\% \times 0.5 = 1.7\%$. Notably, the estimated $\theta$ is of the same order of magnitude as micro-structure estimates of price impact (see e.g. \cite{toth2011anomalous, frazzini2018trading} and \cite{bouchaud2018trades} for more references). Lastly, specification (3) removes the fund-time fixed effect. The coefficient doubles in size suggesting that flows do indeed affect all stocks in a funds portfolio irrespective of the illiquidity of the stock-level positions. It is therefore important to control for fund-time fixed effects, as one cannot rule out that flows contain information about the funds' underlying holdings. Note that the return $r_{n,t}$ on stock $n$ at time $t$ does not vary across investors. The advantage of the pooled triple difference estimator therefore comes at the cost of potentially overestimating (underestimating) $\theta$ if an individual stock receives (offsetting) flow-driven trades of many funds on the same day. In Figure \ref{FIT} we aggregate flow-driven trades at the stock level (FIT) as in \cite{lou2012flow} and find a larger price impact of 0.55 (with a t-Statistic of 16). Therefore, overlapping trades do not seem to inflate the coefficient. When we remove co-held stocks in the estimation of (\ref{Estimation Equation}) the coefficient remains unchanged. Appendix \ref{Ordinary FIT} spells out the difference between FIT and our specification in greater detail. 

\noindent \textbf{Triple Difference Estimator}.
Holding portfolio concentration $\mathcal{C}_{i,n,t}$ constant, a larger fund implies a larger $\mathcal{I}_{i,n,t}$. Holdings size $\mathcal{S}_{i,t}$ constant, a higher position concentration implies a larger $\mathcal{I}_{i,n,t}$. Hence, two concerns that may remain are i) that flows may become more informed as a fund grows in size or ii) that flows are more informed for funds that hold more concentrated positions. In both cases $f_{i,t+1}^{\eta}\mathcal{I}_{i,n,t}$ would be correlated with the error term. To address this concern we note that $f_{i,t+1}^{\eta}\mathcal{I}_{i,n,t}=f_{i,t+1}^{\eta}\mathcal{C}_{i,n,t}\mathcal{S}_{i,t}$. We can therefore separately control for $f_{i,t+1}^{\eta}\mathcal{C}_{i,n,t}$. $f_{i,t+1}^{\eta}\mathcal{S}_{i,t}$ is subsumed by the fund-time fixed effect. This effectively turns $\theta$ into a triple difference estimator that accounts for the possibility that the information in flows is correlated with fund size and position concentration. Specification (3) in Table \ref{OLS Regression Results} reports the results. The triple difference specification leaves the estimated price impact unchanged, suggesting that the information contained in flows is largely unrelated to fund size and position concentration. Appendix \ref{Additional Figures} reports further robustness tests including a horserace of the linear against the square root specification, scaling by volume versus shares outstanding, and including versus excluding the volatility prefactor. We find that at a daily frequency, the square root specification with scaling by volume dominates the standard linear specification in terms of statistical significance and incremental explained variation. This holds both for disaggregated specification (\ref{Estimation Equation}) as well as for the standard FIT specification.

\subsection{Estimating Impact at the Fund Level}
Note, that we can also estimate $\theta$ in a fund-level regression of ETF returns onto ETF flows interacted with fund-level illiquidity as suggested by equation (\ref{Self-Inflated Return Equation}). While aggregating stock-level trades at the fund level removes heterogeneity, it is a more direct test of the hypothesis that fund-level flows affect the returns of illiquid funds (see equation \ref{Self-Inflated Return Equation}). Furthermore, if markets accommodate the flow-driven trades primarily by substituting towards similar stocks held in the respective ETF's portfolio, the price impact may be larger at the fund than at the individual stock level. In other words, groups of stocks (e.g. factors) may have fewer substitutes than individual stocks. \cite{chaudhary2022corporate} provide compelling evidence for lower substitutability across aggregated corporate bond portfolios than for individual bonds within e.g. rating categories. \cite{li2022prices} directly estimates price impact in equities at different levels of aggregation and finds higher multipliers for aggregated portfolios of similar stocks. 

We first decompose aggregate fund returns $R_{i,t}$ into a flow-driven component and a residual component $R^{\perp}_{t}=\sum_n w_{t-1,n}r^{\perp}_{n,t}$ capturing fundamental news or risk factor exposures. Formally, $R_{n,t} = \theta f^\eta_{t} \mathcal{I}_{t-1} + R^{\perp}_{t}$. We can estimate price impact in a pooled regression across fund-time observations, which is simply a portfolio-weighted version of (\ref{Estimation Equation}). We replace position-level returns $r_{n,t}$ and illiquidity $\mathcal{I}_{i,n,t-1}$ with their portfolio-level counterparts $R_{i,t}$ and $\mathcal{I}_{i,t-1}$:
\begin{equation}
\label{Estimation Equation 2}
    R_{i,t} = \alpha_{i} + \alpha_t + \gamma f^\eta_{i,t}   + \beta \mathcal{I}_{i,t-1} + \theta f^\eta_{i,t} \mathcal{I}_{i,t-1} + \text{Controls} + \epsilon_{i,t}
\end{equation}
Note, that as the estimation is at the fund-time level, we cannot control for fund-time fixed effects. We instead control for flows $f^\eta_{i,t}$, fund and time fixed effects separately. Table \ref{OLS Regression Results Fund Level} reports the results.
\begin{table}[H]
\center
\caption{\textbf{Price Impact at the Fund-level}. The table reports the estimated price impact coefficients from pooled OLS regressions using daily data on ETF returns, ETF flows $f^{\eta}_{i,t}$, and fund-level illiquidity $\mathcal{I}_{i,t-1}$. Specification (1) regresses daily fund returns onto flows. Specification (2) estimates the price impact using the interaction of illiquidity and flows $f^{\eta}_{i,t}\mathcal{I}_{i,t-1}$ including fund and time fixed effects. Specification (3) controls for $\mathcal{C}_{i,t-1}$ and the interaction of concentration and flows $f^{\eta}_{i,t}\mathcal{C}_{i,t-1}$. Specification (4) to (6) include fixed effects. In all specifications, we have fixed $\eta=0.5$ such that $f^{\eta}_{i,t}$ are square-root flows. For simplicity of exposition the table reports $f^{\eta}_{i,t}$ and $\mathcal{I}_{i,t-1}$ as $f$ and $\mathcal{I}$ respectively. Standard errors are double clustered at the day, and the stock-day level (to account for the fact the same stock-day observations appear for many investors).}
\begin{tabular}{lllllll}
\toprule
{} &       (1)       &       (2)       &       (3)       &       (4)       &       (5)       &       (6)       \\
                        &                 &                 &                 &                 &                 &                 \\
\midrule
const                   &  -0.0006***     &  -0.0006***     &  -0.0006***     &  -0.0006***     &  -0.0006***     &  -0.0006***     \\
                        &  (-8.0613)      &  (-8.5812)      &  (-8.6846)      &  (-5.0616)      &  (-27.368)      &  (-7.3679)      \\
$f$                     &  0.0020**       &  -0.0023**      &  -0.0018*       &  -0.0018*       &  -0.0020*       &  -0.0020*       \\
                        &  (2.0361)       &  (-2.3702)      &  (-1.7191)      &  (-1.7210)      &  (-1.9272)      &  (-1.9404)      \\
$f\times \mathcal{I}$    &                 &  0.6183***      &  0.7978***      &  0.8073***      &  0.7654***      &  0.7765***      \\
                        &                 &  (7.0562)       &  (5.2081)       &  (5.2604)       &  (5.1069)       &  (5.1699)       \\
$ \mathcal{I}$           &                 &  -0.0071**      &  -0.0072**      &  -0.0123        &  -0.0063**      &  -0.0056        \\
                        &                 &  (-2.0683)      &  (-2.0752)      &  (-0.7963)      &  (-2.1367)      &  (-0.4796)      \\
mkt                     &                 &                 &  0.0245         &  0.0244         &                 &                 \\
                        &                 &                 &  (1.3143)       &  (1.3086)       &                 &                 \\

\midrule
Effects                 &                 &                 &                 &  Entity         &  Time           &  Entity         \\
                        
                    &                 &                 &                 &                 &                 &  Time           \\
Triple Diff.            &                 &                 &           Yes      &  Yes & Yes      &  Yes         \\
No. Observations        &  1175487        &  1175487        &  1175487        &  1175487        &  1175487        &  1175487        \\
R-squared               &  6.952e-05      &  0.0008         &  0.0013         &  0.0013         &  0.0008         &  0.0008         \\
\bottomrule
\end{tabular}

\label{OLS Regression Results Fund Level}
\end{table}
Across all specifications, we find a price impact of around 0.78. Despite the fund-level aggregation, the price impact coefficients are still relatively precisely estimated with t-statistics between 5 and 7. The magnitudes are in line with the stock-level regression without fund-time fixed effects. This underlines the notion that at the portfolio level, price impact can be larger. Funds hold similar stocks in a portfolio, which -- among themselves -- are closer substitutes than aggregate groups of stocks. The fund-time fixed effects in the stock-level specification controls for the higher substitutability among co-held stocks as in \cite{chaudhary2022corporate}.\footnote{For a microstructural interpretation of such ``cross-impact'' effects, see \cite{benzaquen2017dissecting, tomas2022build}.} As we are concerned with the return chasing at the aggregate fund level, we henceforth use the fund-level estimates as the relevant price impact in our structural specifications. However, all results remain qualitatively unchanged (with slightly lower magnitudes) when choosing the conservative stock-level price impact of $\hat{\theta} = 0.33$. Overall, we view the stock-level price impact of 0.78 as a conservative estimate that lies on the lower spectrum of price impact documented by other high-frequency studies that find an impact closer to 1 (\cite{chacko2008price}). One possible reason is that the flow-driven trades combined with the triple difference estimator contain less information than the raw orders typically used in high-frequency studies leading to lower impact estimates.

\subsection{Modelling Reversal}
\begin{figure}[H]
    \centering
\caption{\textbf{Reversal}. The figure plots the cumulative sum of the coefficients from the distributed lag model of daily fund returns onto $S=40$ lags of flow-driven trades $f^\eta_{i,t-s} \mathcal{I}_{i,t-1-s}$ of them form $R_{i,t} = \sum_{s=0}^S\theta_s (f^\eta_{i,t-s} \mathcal{I}_{i,t-1-s})+ \text{Controls} + \epsilon_{i,t}$. The control variables include lagged flows, returns, $\mathcal{C}_{i,t-1-s}$ and $f^\eta_{i,t-s}\mathcal{C}_{i,t-1-s}$  up to 40 days, as well as fund-time fixed effects. The red line reports the simulated estimates of an exponential decay model of the form, $R_{i,t} = \theta_0 f^\eta_{i,t}\mathcal{I}_{i,t-1} + \theta_1 \sum_{s=1}^S e^{-\lambda_{\theta} (s-1)} (f^\eta_{i,t-s} \mathcal{I}_{i,t-1-s})+ \text{Controls} + \epsilon_{i,n,t}$, estimated via nonlinear least squares over the same panel of stock-fund-day observations from 2019 to 2024. The estimated exponential decay coefficients are $(\theta_0, \theta_1, \lambda_{\theta}) = (0.664, -0.087, 0.323)$. The controls are the same as in Table \ref{OLS Regression Results Fund Level} plus daily return lags up to 40 days. The shaded areas indicate 95\% confidence bands where the cumulative standard errors are clustered at the day and stock-day level. Cumulative standard errors are computed accounting for the covariance across coefficient estimates, i.e. $\sqrt{\mathbf{1}_t^T \Omega_t \mathbf{1}_t}$ where $\Omega_t$ is the covariance matrix of the coefficients up until lag $t$ and $\mathbf{1}_t$ is a $t \times 1$ vector of ones.}
\includegraphics[scale=1]{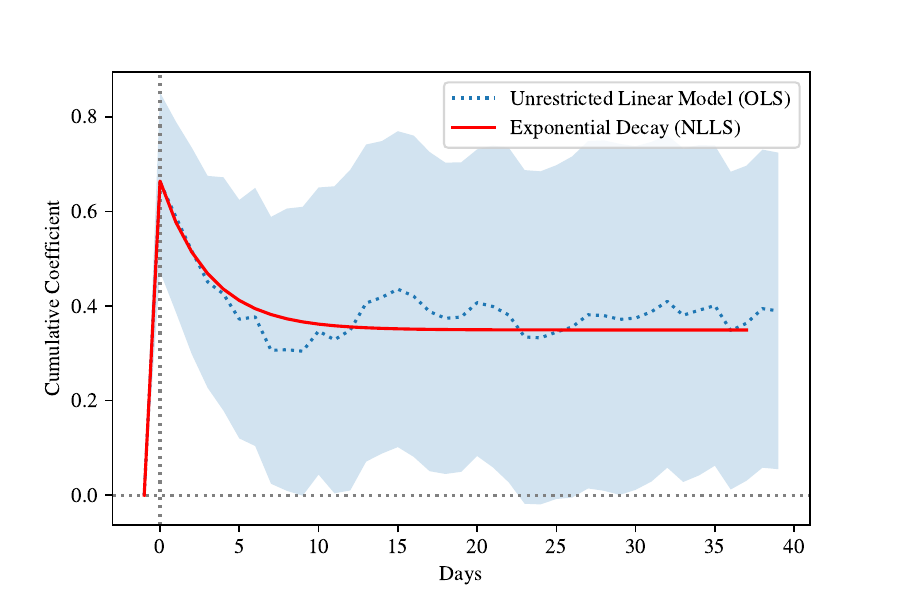}
\label{Reversal}
\end{figure}

Do self-inflated fund returns revert? To this end, it is useful to take a stance on the nature of the price impact of flow-driven trades. At a quarterly frequency (see e.g. \cite{lou2012flow}, \cite{edmans2012real} and \cite{khan2012mutual}), statistically distinguishing between the permanent and transitory impact of flow-driven trading is difficult as the variance of long-run returns greatly exceeds the variance of the initial demand shock. The high-frequency nature of our data allows us to estimate reversal with greater precision. We estimate permanent versus transitory impact in a distributed lag model of daily fund-level returns onto $S=40$ lags of flow-driven trades. Formally, 
\begin{equation}
\label{Impact Reversal}
    R_{i,t} = \sum_{s=0}^S\theta_s (f^\eta_{i,t-s} \mathcal{I}_{i,t-1-s})+ \text{Controls} + \epsilon_{i,t} 
\end{equation}
The long run price impact until $t^*$ is given by the sum of the coefficients $\sum_{s=0}^{t^*}\theta_s$. Figure \ref{Reversal} plots the cumulative sum of the coefficients. \cite{bucci2018slow} (Figure 3) estimate a strikingly similar relaxation of impact for single stocks using institutional trades from Ancerno.
We find that part of the initial price impact reverts over the following 5-10 days leading to a long-run impact of roughly 0.4. We also estimate a parsimonious exponential decay model by parameterizing $\theta_s$ as a function of time. Specifically, we assume an exponentially reverting transient price impact. This implies estimating the level of the impact and the speed of reversion, as opposed to heterogenous coefficients on all the lags. Formally,  
\begin{equation}\label{eq:relax}
    R_{i,t} = \underbrace{\theta_0 f^\eta_{i,t}\mathcal{I}_{i,t-1}}_{\text{Initial Impact}} + \underbrace{\theta_1 \sum_{s=1}^S e^{-\lambda_{\theta} (s-1)} (f^\eta_{i,t-s} \mathcal{I}_{i,t-1-s})}_{\text{Reversal}}+ \text{Controls} + \epsilon_{i,n,t},
\end{equation}
where $\theta_0$ is the contemporaneous price impact, $\theta_1$ measures the impact decay and $\lambda_{\theta}$ measures the speed of the decay.\footnote{There is a subtle point here: we expect the square-root dependence of impact at high frequencies to become linear at low frequencies, whereas Eq. \eqref{eq:relax} assumes a square-root dependence for all frequencies. A better model accounting for such a cross-over is discussed in \cite{bouchaud2022inelastic}, but is cumbersome and more difficult to calibrate.} The long-term price impact is given by $\theta_0-\theta_1/\lambda_{\theta}$. The red line in Figure \ref{Reversal} plots the fitted price impact from the exponential decay model. The nonlinear model fits the cumulative sum of the reduced form coefficients remarkably well. The estimated exponential decay coefficients are $(\theta_0, \theta_1, \lambda_{\theta}) = (0.664, -0.087, 0.323)$. In all subsequent sections, we take this partial reversal of price impact into account whenever we aggregate impact returns over time. See Appendix \ref{Aggregation} for details. 

\subsection{Variance Decomposition of Fund Returns}
To what extent are self-inflated returns economically meaningful relative to the total returns that funds generate? We decompose realized fund returns into a self-inflated component $R^{\mathcal{I}}_{i,t}$ and a residual component $R^{\perp}_{i,t}$ capturing fundamental returns drivers such as risk exposures or news.
\begin{equation*}
    R_{i,t} = \underbrace{\hat{\theta}f^{\eta}_{i,t}\mathcal{I}_{i,t-1}}_{R^{\mathcal{I}}_{i,t}} + R^{\perp}_{i,t}
\end{equation*}
where the self-inflated component is computed using the estimated price impact $\hat{\theta}=0.78$ from the fund-level difference-in-difference estimator (Table \ref{OLS Regression Results Fund Level}, model (6)). When aggregating returns at different frequencies, we use the coefficients from the exponential decay model estimated above.

In most neoclassical models, price impact ($\theta$) is close to 0, and therefore flows do not affect realized returns ($R_{i,t} \approx R^{\perp}_{i,t}$). In the presence of non-zero price impact $\theta>0$ returns are driven by both fundamentals and price impact. The impact depends on the liquidity of the underlying portfolio $\mathcal{I}_{i,t}$ and on the size of the flows $f_{i,t}$. Even if $\theta$ is large, $R_{i,t}^{\mathcal{I}}$ is still close to zero for most ETFs that track a broad index because either $\mathcal{I}_{i,t-1}$ or $f_{i,t}$ are small. The extent to which self-inflated returns are economically meaningful relative to total returns is therefore an empirical question. For each fund $i$, the fraction of total return variation explained by self-inflated returns is given by $\beta_i^{\mathcal{I}} = \frac{cov_i(R^{\mathcal{I}}_{i,t}, R_{i,t})}{var_i(R_{i,t})}$. Note, that self-inflated fund returns $R^{\mathcal{I}}_{t} =  \theta f_t^{\eta}\mathcal{I}_{t-1}$ are large, when funds' illiquidity $\mathcal{I}_{t}$ is high, i.e. when the fund is both large (high $\mathcal{S}$) and holds a concentrated portfolio $\mathcal{I}$. Therefore we expect $\beta_i^{\mathcal{I}}$ to be higher for large concentrated funds (i.e. the funds in the northeast corner of Figure \ref{Concentration versus Size}).

\begin{figure}[H]
\centering
\caption{\textbf{Variance Decomposition: Monthly ETF Returns.}. The figure plots the explained variance of monthly fund returns due to self-inflated returns $\beta_i^{\mathcal{I}} = cov_i(R^{\mathcal{I}}_{i,t}, R_{i,t})/var_i(R_{i,t})$. We group funds by fund size $\mathcal{S}$ and portfolio concentration $\mathcal{C}$ and compute the $\beta_i^{\mathcal{I}}$. The horizontal axis displays fund size quintiles. The orange bars represent the top 10\% of funds with the most concentrated portfolios within each size quintile. The blue bars represent the remaining 90\%.}
\center
    \includegraphics[scale=0.9]{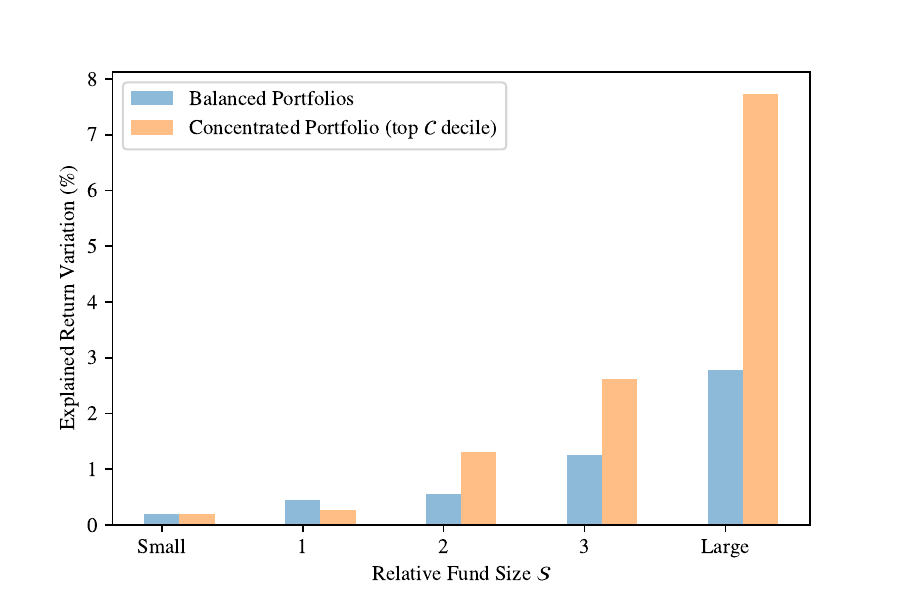}
    \label{Variance Decomposition}
\end{figure}

\section{Do fund flows chase self-inflated returns?}
\label{Impact Chasing Section}
We now investigate, whether fund flows chase self-inflated returns at a high frequency. The flow-performance relationship for mutual funds is typically tested by regressing quarterly fund flows on past long-run returns.\footnote{See e.g. \cite{ippolito1992consumer}, \cite{chevalier1997risk}, \cite{huang2007participation}, and \cite{goldstein2017investor}.} \cite{barber2016factors} and \cite{dannhauser2019flow} regress monthly flows onto monthly return lags and find that the first lag contributes most strongly to the aggregate flow-performance relationship. It is possible that the self-inflated return only materially affects overall returns over an even shorter window. This can be either because the price impact quickly reverts or because fundamental returns are volatile, which conceals the self-inflated return at longer horizons. 
If investors primarily allocate flows based on average long-run returns, the positive feedback loop from self-inflated returns is potentially small. 

\subsection{High-frequency Return Chasing}
We therefore first test the origins of the low-frequency flow-performance relationship. To what extent is the flow-performance relationship driven by the most recent returns? If investors place a higher weight on the most recent fund return, even a quickly dissipating price impact can affect future inflows. Following \cite{schmickler2020identifying}, we regress daily ETF flows on up to $L=200$ lags of daily ETF returns and flows. In essence, we estimate the weights that investors place on lagged returns at a daily frequency. 
\begin{equation}
\label{High Frequency Return Chasing}
    f_{i,t+1} = \alpha_t + \sum_{s=0}^{L} \beta_s R_{i,t-s} + \text{Controls} + \epsilon_{i,t+1}
\end{equation}
The return chasing kernel, $\{\beta_s\}_{s=0}^L$ is the weighting scheme that investors place on past returns when allocating future flows. We estimate the return chasing kernel over the panel of ETF flows and returns from 2019 to 2024 and report the cumulative sum of the coefficients over increasing time lags. Figure \label{Return Chasing Kernel} plots the results. 

\begin{figure}[h]
\centering
\center
\caption{\textbf{Return Chasing Kernel}. The figure plots the cumulative return-chasing coefficient from the following model $f_{i,t+1} = \alpha_t + \sum_{s=0}^{L} \beta_s R_{i,t-s} + \text{Controls} + \epsilon_{i,t+1}$. The blue line reports the cumulative sum of the OLS coefficients $\{\sum_{s=0}^k\beta_s\}_{k=1}^{L}$ using $L=200$ lags. The shaded areas indicate 95\% confidence bands where the cumulative standard errors are clustered at the day and fund level. Cumulative standard errors are computed accounting for the covariance across coefficient estimates, i.e. $\sqrt{\mathbf{1}_t^T \Omega_t \mathbf{1}_t}$ where $\Omega_t$ is the covariance matrix of the coefficients up until lag $t$ and $\mathbf{1}_t$ is a $t \times 1$ vector of ones. The red line reports the cumulative return chasing from an exponential decay model estimated: $f_{i,t+1} = \beta\sum_{s=0}^{65} e^{-\lambda_{\beta} s} R_{i,t-s} + \text{Controls} + \epsilon_{i,t+1}$. We estimate $\hat{\lambda}_{\beta}=0.01$ via nonlinear least squares (NLLS) over the panel of daily ETF-flow observations from 2019 to 2024.}
  \includegraphics[]{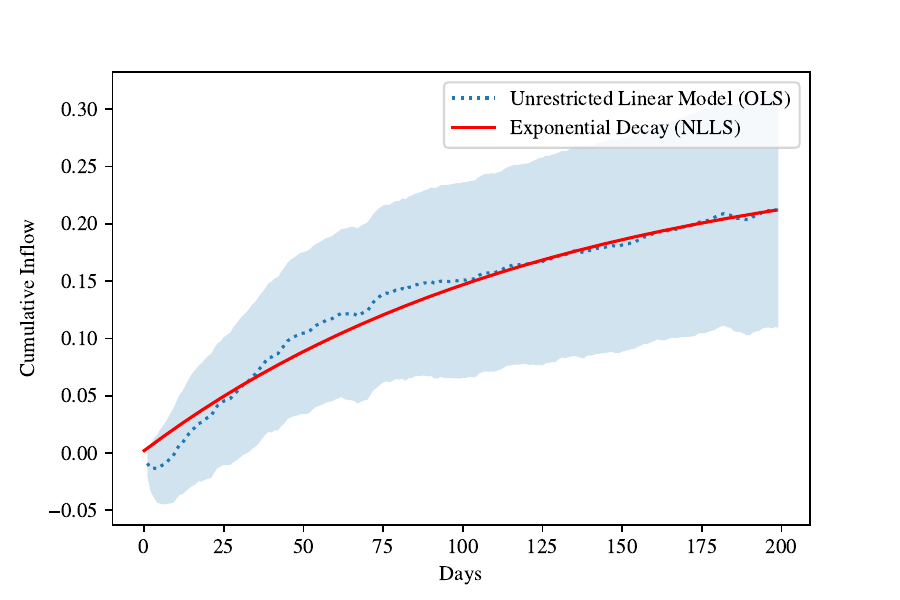}  
\label{Return Chasing Kernel}
\end{figure}

The figure shows that more recent returns receive a higher weight in the return chasing kernel. Thus the flow-performance relationship is predominantly driven by the most recent returns. This is in line with exponential decaying weights on past returns in extrapolative expectations (see \cite{greenwood2014expectations}, \cite{barberis2015x}, and \cite{da2021extrapolative}). Following \cite{barber2016factors} we also estimate an exponential decay model (red line) which well-fits the cumulative sum of the OLS coefficients. The exponential decay parameterizes the coefficients on all the lags as a function of time, $\beta_s=\beta e^{-\lambda_{\beta} s}$. We estimate $\hat{\lambda}_{\beta}=0.01$ via nonlinear least squares (NLLS) over the panel of daily ETF-flow observations from 2019 to 2024. The exponential decay formulation allows for a parsimonious representation of the return-chasing behavior that is unrelated to the horizon. This is useful when comparing the return-chasing behavior across different return decompositions, such as price impact versus fundamental return (as in this paper), or abnormal versus factor return (as in \cite{barber2016factors}). The cumulative coefficient implies that a 10\% return realization leads to an additional $10\% \times 0.2 = 2\%$ inflows, the first 1\% occurring within 50 days of the initial return. In Figure \ref{Return Chasing Splits} we show that the exponential weighting scheme holds generally across sample splits, such as for both positive and negative past returns, small and large, liquid and illiquid ETFs, and when estimating the coefficients via AUM-weighted least squares. Notably, we find a stronger flow-performance sensitivity when we restrict the sample to active funds. The exponential over-weighting of the most recent return is what allows flow-driven price impact to affect the cross-section of fund flows. If investors weighted all past return realizations equally, then self-inflated returns would have a smaller effect on flows: Over long time scales, average returns are less affected by self-inflated returns which are primarily driven by short-lived spikes in inflows. However, because investors place a high weight on the most recent return, even short-lived self-inflated returns can strongly affect future flows leading to a positive feedback loop. The strength of the effect depends on i) how much of the variation in fund returns is driven by self-inflated returns and ii) how persistent the price impact is.

\subsection{Chasing Impact versus Fundamentals}
\label{Chasing Impact versus Fundamentals}
We now decompose realized fund returns into a self-inflated component (driven by flow-induced trades) and a residual (fundamental) component, taking into account price impact reversal. To this end, note that a fund's returns are not only impacted by its self-inflated return but also by the self-inflated return of all other funds that hold overlapping securities. The residual component $R^{\perp}_{i,t}$ is the remaining fund return that is unexplained by the aggregate self-inflated returns.
\begin{equation}
    R_{i,t} = R^{\mathcal{I}}_{i,t} + R^{\perp}_{i,t}
\end{equation}
where, with a slight abuse of notation, $R^{\mathcal{I}}_{i,t}$ measures the price impact from flow-driven trades of \textit{all} funds including the reversal from past trades. Appendix \ref{Aggregation} provides further details on how $R^{\mathcal{I}}_{i,t}$ is constructed. Following the methodology proposed by \cite{barber2016factors} we construct a weighted average of past returns based on the exponential decay model estimated above. In particular, we define weighted versions of past fundamental and impact returns as $\tilde{R}^{\perp}_{i,t} = \sum_{s=0}^{L} w_s R^{\perp}_{i,t-s}$ and $\tilde{R}^{\mathcal{I}}_{i,t} = \sum_{s=0}^{L} w_s R^{\mathcal{I}}_{i,t-l}$ where the weights $w_s = \frac{e^{-\lambda_{\beta} s}}{\sum^{L}_{s=0} e^{-\lambda_{\beta} s}}$ are determined by the exponential decay $\lambda_{\beta}$ of the flow-return sensitivity. Regressing daily flows on the exponentially weighted past return allows comparing individual coefficient sensitivity coefficients as opposed to 200 coefficients on the different lags. The decomposition $\tilde{R}_{i,t}=\tilde{R}^{\mathcal{I}}_{i,t} + \tilde{R}^{\perp}_{i,t}$ allows assessing whether investors equally chase both impact returns ($\tilde{R}^{\mathcal{I}}_{i,t}$) and fundamental returns ($\tilde{R}^{\perp}_{i,t}$) when deciding how to allocate their capital. We run the decomposed regression
\begin{equation}
\label{Decomposed Specification}
     f_{i,t+1} = \alpha_t +  \beta_{1}\tilde{R}^{\mathcal{I}}_{i,t} + 
     \beta_{2} \tilde{R}^{\perp}_{i,t} + Controls + \epsilon_{i,t+1}
\end{equation}
and report the results in Table \ref{Impact Chasing}. 
\begin{table}[h]
\center
\caption{\textbf{Price Impact Chasing}. The table reports the estimated coefficients of pooled OLS regressions of daily flows onto (exponentially weighted) average past returns, as well as decomposed average returns into an impact and a fundamental component. Specification (1) reports the coefficient to the regression on raw returns $f_{i,t+1} = \alpha_t +  \beta\tilde{R}+ Controls + \epsilon_{i,t+1}$. Specification (2) reports the coefficients to the regression on the decomposed returns $f_{i,t+1} = \alpha_t +  \beta_{1}\tilde{R}^{\mathcal{I}}_{i,t} + 
     \beta_{2} \tilde{R}^{\perp}_{i,t} + Controls + \epsilon_{i,t+1}$. All specifications control for a constant, lagged flows up to 200 days, and time-fixed effects. The sample period is 2019 to 2024. Standard errors are double clustered at the day and the fund level.}
\begin{tabular}{lllll}
\toprule
{} &       (1)       &        (2)       \\
                                &                 &                  \\
\midrule
$R$                   &  0.2136***      &                  \\
                                &  (3.4301)       &                  \\                                
$R^{ \mathcal{I}}$     &                 &  0.2375***       \\
                                &                 &  (3.5943)        \\
$R^{\perp}$           &                 &  0.2229***       \\
                                &                 &  (3.3425)        \\
\midrule
Controls                         &  Yes           &  Yes            \\
Effects                         &  Time           &  Time            \\
No. Obs.                &  1139565        &  1139565         \\
R-squared                       &  0.0150         &  0.0150          \\
\bottomrule
\end{tabular}

\label{Impact Chasing}
\end{table}
To set the stage we first regress daily fund flows onto undecomposed average past returns $\tilde{R}_{i,t}$ and obtain a coefficient of 0.21 (t-stat 3.4).\footnote{Note, that this is roughly equal to the sum of all coefficients of regression (\ref{High Frequency Return Chasing}).} We then regress daily fund flows onto decomposed returns $\tilde{R}^{\mathcal{I}}_{i,t}$ and $\tilde{R}^{\perp}_{i,t}$ according to (\ref{Decomposed Specification}). If investors are able to differentiate between price impact and fundamental returns, we should observe that the coefficient on $\tilde{R}^{\mathcal{I}}_{i,t}$ is 0 and statistically insignificant as a high price impact is not reflecting higher future expected returns. However (and perhaps unsurprisingly), we find that fund flows cannot distinguish the two sources of past returns and significantly respond to both components. We furthermore cannot reject the null hypothesis that the coefficients are identical. When observing past performance, investors are hence not able to differentiate between price impact and fundamental returns (or `managerial skill'). In Appendix Table \ref{Impact Chasing AP} we split the sample into active (non-benchmarked) and passive (benchmarked) funds. For passive funds, the coefficients on $\tilde{R}^{\mathcal{I}}_{i,t}$ and $\tilde{R}^{\perp}_{i,t}$ are 0.19 (t-stat 2.9) and 0.18 (t-stat 2.7) respectively. For active funds, the coefficients are 0.77 (t-stat 4.3) and 0.89 (t-stat 8.1) respectively. For both active and passive funds we can strongly reject the null hypothesis that investors do not pay attention to self-inflated returns. We furthermore cannot reject that the return-chasing coefficient is the same for both self-inflated and fundamental returns.

\section{Self-Inflated Feedback Loops}
\label{Self-Inflated Feedback Loops}
The previous section showed that because investors place a higher weight on the most recent return, even short-lived price pressure can have an impact on the distribution of fund flows. This can cause an endogenous feedback loop: Flows cause a price impact, which is a realized return that causes further inflows and amplifies the initial price impact. The magnitude of this feedback loop remains an empirical question, which we address in this section.

\subsection{Ponzi Flows}
\begin{figure}[h]
\centering
\center
\caption{\textbf{Ponzi Flows Relative to Total Flows.} The figure plots the fraction of fund flow volume that can be attributed to chasing past price impact. Formally, we compute $\frac{\sum_i |f^{\mathcal{P}}_{i,t+1}|A_{i,t}}{\sum_i |f_{i,t+1}|A_{i,t}}$ over different subsets of funds and report the rolling monthly mean. The red line reports the fraction for the top decile of illiquid funds (highest $\mathcal{I}$). The black line reports the fraction for all other funds.}
\includegraphics[scale=0.7]{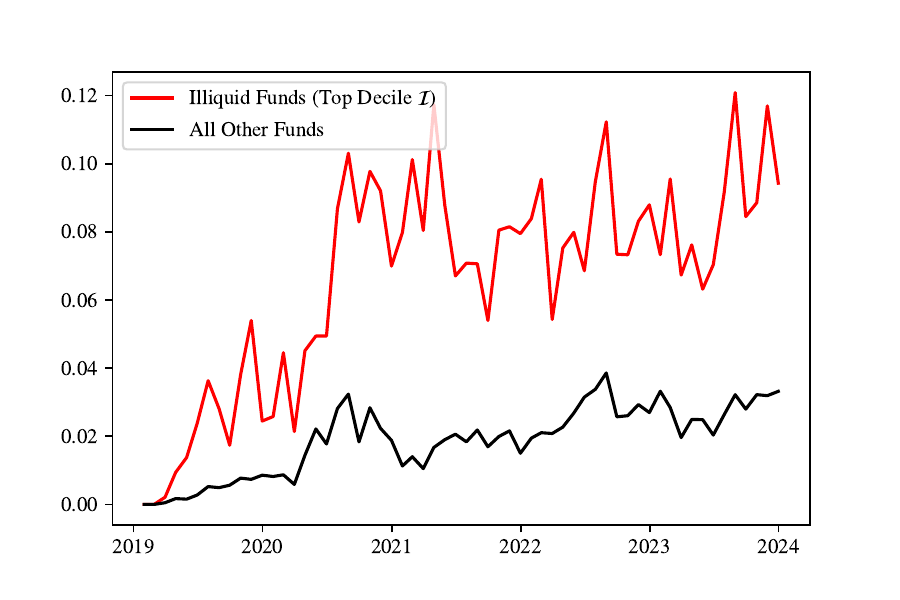}
\label{Ponzi Flows}
\end{figure}

We follow \cite{darmouni2022nonbank} and use the linear flow-performance relationship as a \textit{structural} equation.\footnote{\cite{darmouni2022nonbank} microfound the linear relationship from households' logit demand over the cross-section of fund characteristics.} Combined with the exponentially decaying price impact, this allows us to decompose fund flows into a self-inflated and a fundamental component, identify bubble funds (i.e. funds with strong self-inflated returns), and predict drawdowns. We define Ponzi flows $f^{\mathcal{P}}_{i,t+1}$ as the component of flows that chase past price impact,
\begin{equation}
    f^{\mathcal{P}}_{i,t+1} \equiv \hat{\beta_{1}}\tilde{R}^{\mathcal{I}}_{i,t},
\end{equation}
where $\hat{\beta_{1}}$ is the estimated coefficient from equation (\ref{Decomposed Specification}). Note, that $\tilde{R}^{\mathcal{I}}_{i,t}$ measures the price impact on $i$'s portfolio from the flow-driven trades of \textit{all} funds and includes the reversal from previous trades.\footnote{See Appendix \ref{Aggregation} for details.} To illustrate the economic importance of price impact chasing flows, we compute the fraction of total flow volume that can be attributed to $f^{\mathcal{P}}_{i,t+1}$. Formally, we compute $\frac{\sum_i |f^{\mathcal{P}}_{i,t+1}|A_{i,t}}{\sum_i |f_{i,t+1}|A_{i,t}}$. Figure
 \ref{Ponzi Flows} plots the relative Ponzi volume over time.
Among the most illiquid funds (top decile $\mathcal{I}$), around 10\% of the daily flow volume can be attributed to Ponzi flows, i.e. flows chasing past price impact. Among all other funds, Ponzi flows are still sizeable and account for 2-3\% of total flow volume.


\subsection{Identifying Bubble Funds}
\begin{figure}[H]
\centering
\center
\caption{\textbf{ETF Bubbles and Ponzi Flows}. The figure plots the cumulative return of all ETFs that outperformed the market by over 50\% in the past two years. The black line reports the average cumulative return across ETFs over the event window. The red line plots the cumulative return of the subset of bubble ETFs whose cumulative Ponzi flows are in the top 30\%, 20\%, 10\% and 5\% among run-up ETFs.}
\begin{subfigure}{0.48\textwidth}
\centering
\includegraphics[scale=0.5]{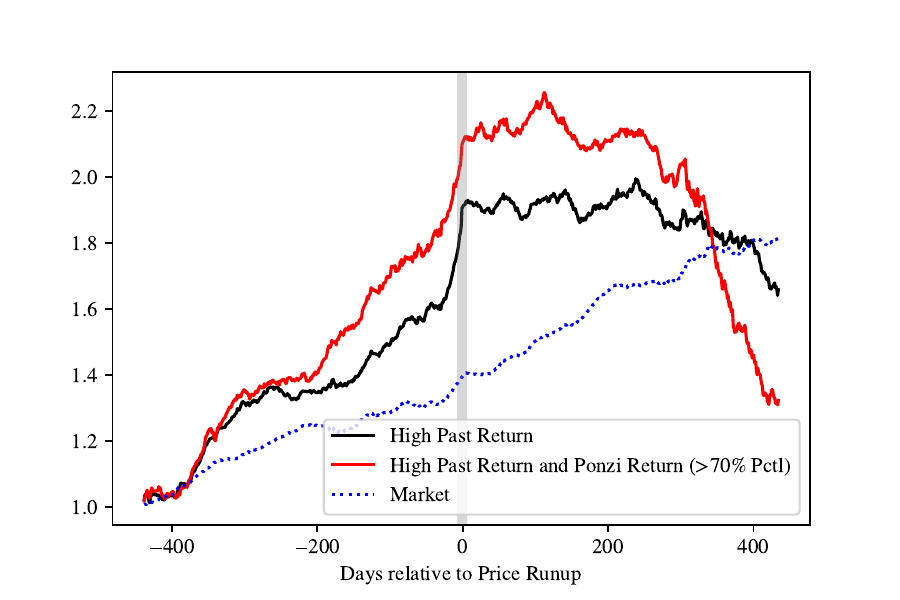}
\caption{Top 30\% cumulative $f^{\mathcal{P}}_{i,t}$}
\end{subfigure}
\begin{subfigure}{0.48\textwidth}
\centering
\includegraphics[scale=0.5]{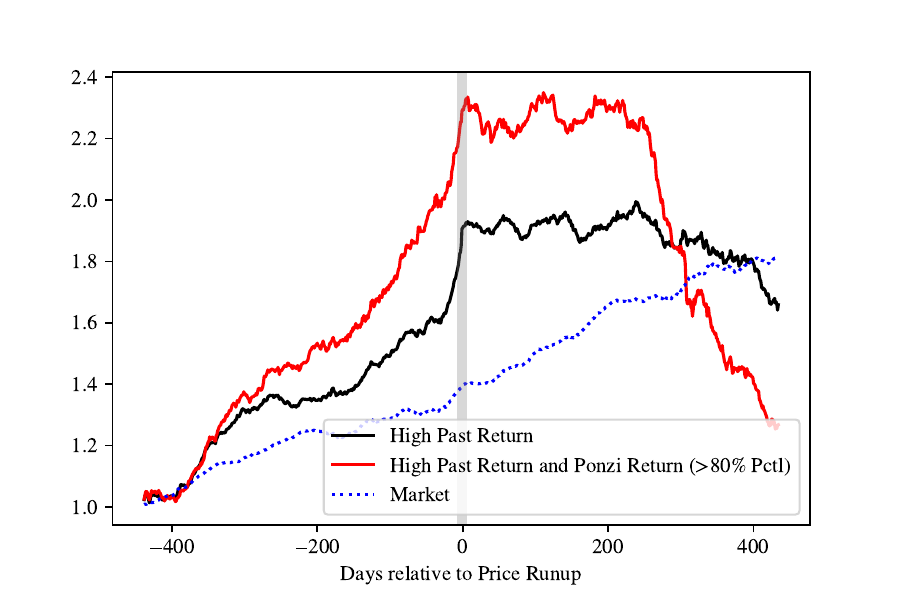}
\caption{Top 20\% cumulative $f^{\mathcal{P}}_{i,t}$}
\end{subfigure}
\begin{subfigure}{0.48\textwidth}
\centering
\includegraphics[scale=0.5]{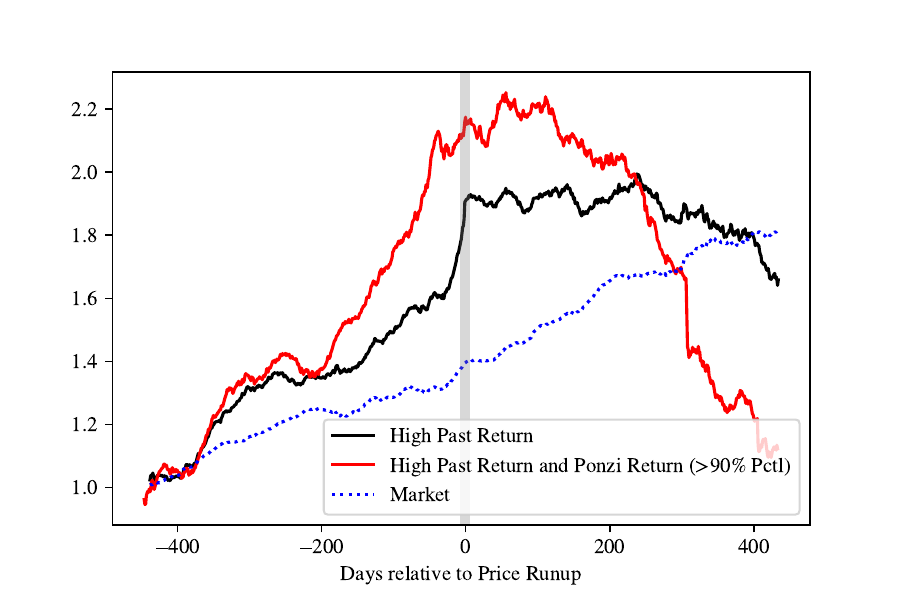}
\caption{Top 10\% cumulative $f^{\mathcal{P}}_{i,t}$}
\end{subfigure}
\begin{subfigure}{0.48\textwidth}
\centering
\includegraphics[scale=0.5]{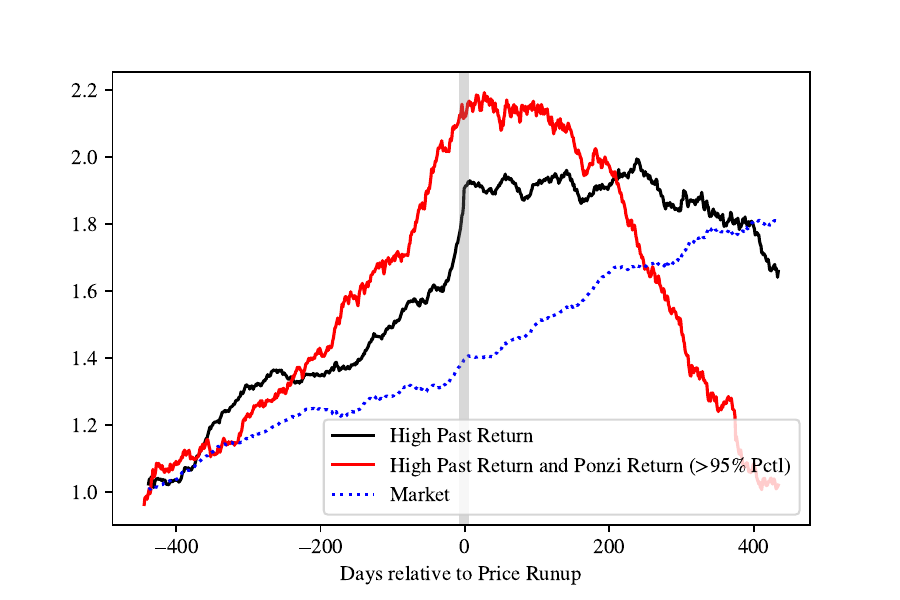}
\caption{Top 5\% cumulative $f^{\mathcal{P}}_{i,t}$}
\end{subfigure}
\label{Bubbles}
\end{figure}

Following \cite{greenwood2019bubbles}, we investigate whether flow-driven trading predicts crashes of ETFs with extreme price runups. To this end, we select `run-up ETFs' that outperformed the market by over 50\% over the previous two years at any point in our sample. We compute the cumulative future 1-year returns for all run-up ETFs as well as for `bubble ETFs', whose cumulative Ponzi flow $f^{\mathcal{P}}_{i,t}$ is in the top 30\%, 20\%, 10\%, 5\% among run-up funds. Figure \ref{Bubbles} plots the cumulative returns over the event window for the run-up ETFs and the bubble ETFs. 

The black line reports the average cumulative return over the event window across all run-up ETFs. As in \cite{greenwood2019bubbles}, excessive outperformance is on average not followed by a subsequent crash. Cumulative returns rather converge back to the market return. The red line splits the sample of run-up ETFs into the funds that received the highest cumulative Ponzi flows $f^{\mathcal{P}}_{i,t}$ during the period. On average, these bubble funds experienced steep crashes, with cumulative returns exceeding -200\% in the two years following the runup. However, as in \cite{greenwood2019bubbles}, timing the crash is difficult. On average, bubble ETFs do not crash within the first year of the run-up.

\subsection{Ponzi Returns}
\begin{figure}[H]
    \centering
\center
\caption{\textbf{Ponzi Wealth Reallocation}. The figure plots the daily dollar amount of capital reallocated via Ponzi flows (left vertical axis) as well as the cumulative sum (right vertical axis). We define Ponzi returns $R_{i,t}^{\mathcal{P}}$ as the return that is caused by flows chasing past price impact. Weighting by lagged assets under management $A_{i,t-1}$ and summing across funds yields the total dollar wealth reallocation due to Ponzi flows $\sum_i R_{i,t}^{\mathcal{P}}A_{i,t-1}$. }
\centering
\includegraphics[scale=0.7]{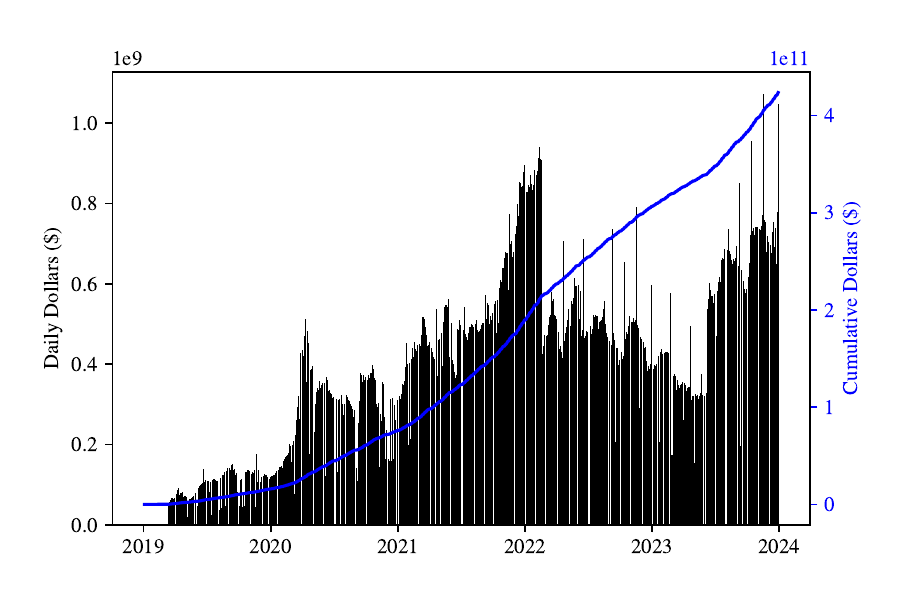}
\label{AUM Ponzi}
\end{figure}

Can Ponzi flows, i.e. flows chasing past price impact, meaningfully affect asset prices? We now turn to the wealth reallocation that is caused by the price impact of Ponzi flows. To this end, we define Ponzi returns as the price impact caused by reinvested Ponzi flows $R^{\mathcal{P}}_{i,t} \equiv \theta \mathcal{I}_{i,t-1}f^{\mathcal{P}}_{i,t}$. The daily dollar reallocation of wealth due to Ponzi return is given by $\sum_i |R^{\mathcal{P}}_{i,t}|A_{i,t-1}$. Figure \ref{AUM Ponzi} plots the total wealth reallocation due to Ponzi returns from 2019 to 2024. 
Every day, ETFs reallocate around \$500 million among investors because of the price impact from flows that chase past price impact. Since 2019, the total cumulative dollars that ETFs have reallocated via this endogenous feedback loop is around 440 Billion.

\section{Conclusion}
\label{Conclusion}

Price impact is a realized return for the incumbent holders of an asset. The inability of market participants to distinguish between realized fund returns stemming from fundamental determinants and those driven by price impact has important implications for asset markets. We show that among ETFs alone, \$500 Million dollars are reallocated daily because market participants chase the price impact of past flow-driven trades. 

The focus of this paper lies on managed funds and the self-inflated feedback loop that arises from the combination of flow-driven trading and the flow-performance relationship. ETFs are a suitable setting to establish first evidence for the interplay of price impact and belief formation because i) we can directly observe demand (via flows) and ii) the proportional reinvestment of flows at a daily frequency allows for cleanly identifying the price impact of non-fundamental demand shocks. 

However, the underlying drivers of this feedback loop -- return chasing and price impact -- apply more generally. For example, trend following in futures markets and the \$300 Billion wide CTA industry, are likely prone to self-inflated price spirals (see \cite{lemperiere2014trend} and references therein). Similarly, the ``quant crunch'' in 2007 (\cite{khandani2011happened}) and other deleveraging spirals can be seen as outcomes of the interaction between return chasing and price impact (\cite{brunnermeier2009market, bouchaud2012impact, cont2013running, kyle2023large}). The increased availability of investor-level holdings and flow data allows quantifying these effects. 
Revisiting these studies through the lens of portfolio holdings, for example via structural models of investor demand as in \cite{koijen2019demand}, is an exciting avenue for future research. Quite remarkably, our price impact estimate from flow-induced trades matches well the estimates from the recent microstructure literature (e.g. \cite{toth2011anomalous, frazzini2018trading, bouchaud2018trades}). Bridging the gap between low-frequency portfolio holdings data and trade data at a higher frequency provides an explicit link between market micro-structure, intermediary asset pricing, and (eventually) corporate finance.

\bibliography{Bibliography}
\bibliographystyle{chicago}
\clearpage
\clearpage

\makeatletter
\renewcommand{\thesubsection}{\Alph{section}.\arabic{subsection}}
\renewcommand{\thetable}{\Alph{section}.\arabic{table}}
\renewcommand{\thefigure}{\Alph{section}.\arabic{figure}}
\makeatother
\pagebreak
\begin{appendices}

\section{Aggregating Self-Induced Fund Returns}
\label{Aggregation}

Self-induced returns do not only impact the fund's own realized return but also the realized returns of all funds $i=1,...,I$ that are invested in the affected securities. The return of fund $i$ that is driven by its own flows is given by $R^{ \mathcal{I}}_{i,t} = \theta  \mathcal{I}_{i,t-1}f^\eta_{i,t}$ where $ \mathcal{I}_{i,t-1} = \sum_n w_{i,t-1,n}  \mathcal{I}_{i,t-1,n}$ is portfolio level illiquidity. Flows into fund $i$ also affect the return of other portfolios (here $j$) that hold overlapping stocks. We therefore define the cross-fund illiquidity as
\begin{equation}
     \mathcal{I}_{t,ij} \equiv \sum_{n=1}^N w_{j,t,n} \mathcal{I}_{i,t}
\end{equation}
Note that $ \mathcal{I}_{ij,t} = \mathcal{I}_{i,t}$ if fund $j$ holds the same portfolio weights as fund $i$. Note also, that $ \mathcal{I}_{t,ij} \neq  \mathcal{I}_{t,ji}$ unless both funds hold the same portfolio weights and have the same assets under management. The total flow-induced return for fund $i$ is therefore
\begin{equation}
     R^{ \mathcal{I},\text{Total, Permanent}}_{i,t} \equiv \theta \sum_{j=1}^I  \mathcal{I}_{t-1,ji} f_{j,t}^{\eta}
\end{equation}
where $\sum_j$ is summing over all $I$ funds including $i$. 
Note that this assumes permanent price impact, in that past flows do not affect the current return. Including price impact reversal (see (\ref{Impact Reversal})) yields
\begin{equation}
     R^{ \mathcal{I},\text{Total}}_{i,t}  \equiv  \sum_{s=0}^S \theta_s \sum_{j}  \mathcal{I}_{t-1-s,ji} f_{j,t-s}^{\eta}
\end{equation} 
where $\theta_0$ is the contemporaneous impact and $\{\theta_s\}_{s>0}$ captures the reversal from past flow-driven trades. The flow-performance relationship suggests that the sensitivity of flows to past returns decays at speed $\lambda_{\beta}$. Therefore, long-run average past returns are computed using the exponential $\lambda_{\beta}$-weighted as described in \ref{Chasing Impact versus Fundamentals}. We compute the weighted average past return using total flow-induced trading including reversal $R^{ \mathcal{I},\text{Total}}_{i,t}$ as
\begin{equation}
    \tilde{R}^{ \mathcal{I}}_{i,t} \equiv \sum_{s=0}^{L} w_s R^{ \mathcal{I},\text{Total}}_{i,t} 
\end{equation}
where $w_s = \frac{e^{-\lambda_{\beta} s}}{\sum^{L}_{s=0} e^{-\lambda_{\beta} s}}$ are the exponential decay weights following \cite{barber2016factors}.

\section{Additional Figures}
\label{Additional Figures}

\begin{figure}[H]
\centering
\caption{\textbf{Daily ETF Returns and Inflows: Robustness}. The figure plots the daily average return of ETFs across different flow buckets. Every day, we sort ETFs into deciles based on their inflows and compute their average return. Subfigure (a) and (b) report average excess returns and CAPM alphas based on sorts by flows relative to AUM $f_{i,t+1}=F_{i,t+1}/A_{i,t}$ and fund illiquidity $ \mathcal{I}_{i,t}$. Subfigure (c) reports returns for an alternative sort based on flows relative to underlying liquidity $F_{i,t+1}/V_{i,t}$ and portfolio concentration $\mathcal{C}_{i,t}$. Note, that $F_{i,t+1}V_{i,t}^{-1}\mathcal{C}_{i,t} = f_{i,t+1}  \mathcal{I}_{i,t}$.}
\begin{subfigure}{0.45\textwidth}
    \includegraphics[scale=0.5]{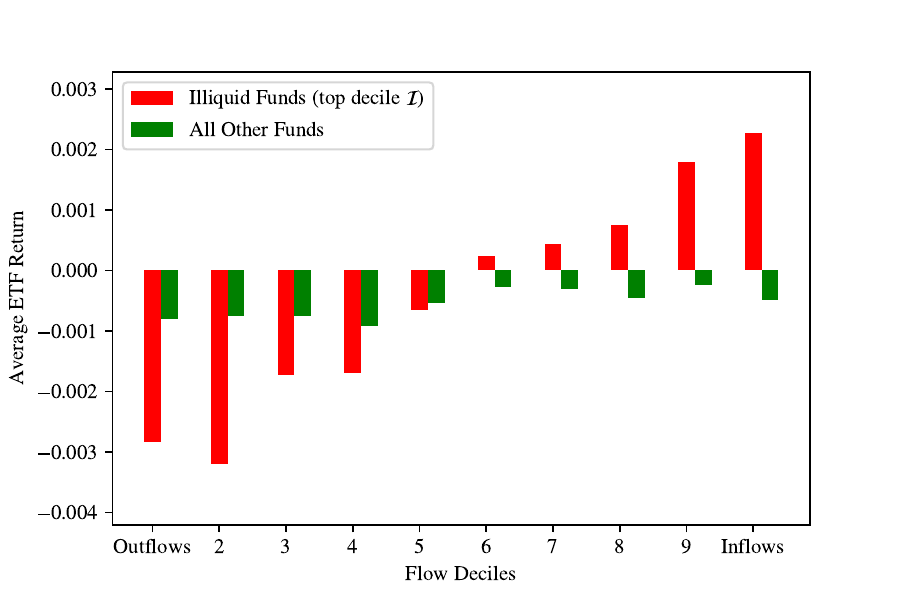}  
    \caption{Excess Returns}
\end{subfigure}
\begin{subfigure}{0.45\textwidth}
    \includegraphics[scale=0.5]{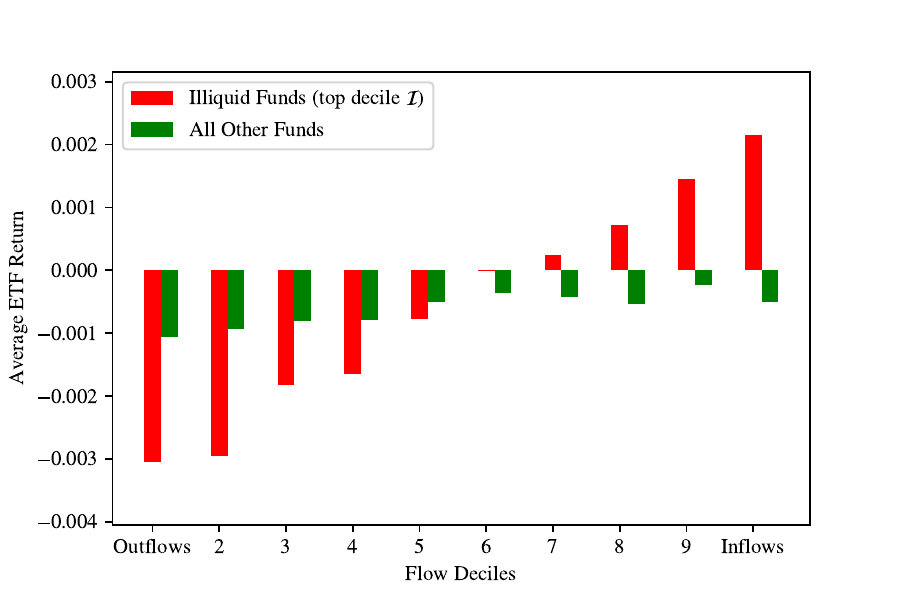}  
    \caption{CAPM Alphas}
\end{subfigure}
\begin{subfigure}{0.45\textwidth}
    \includegraphics[scale=0.5]{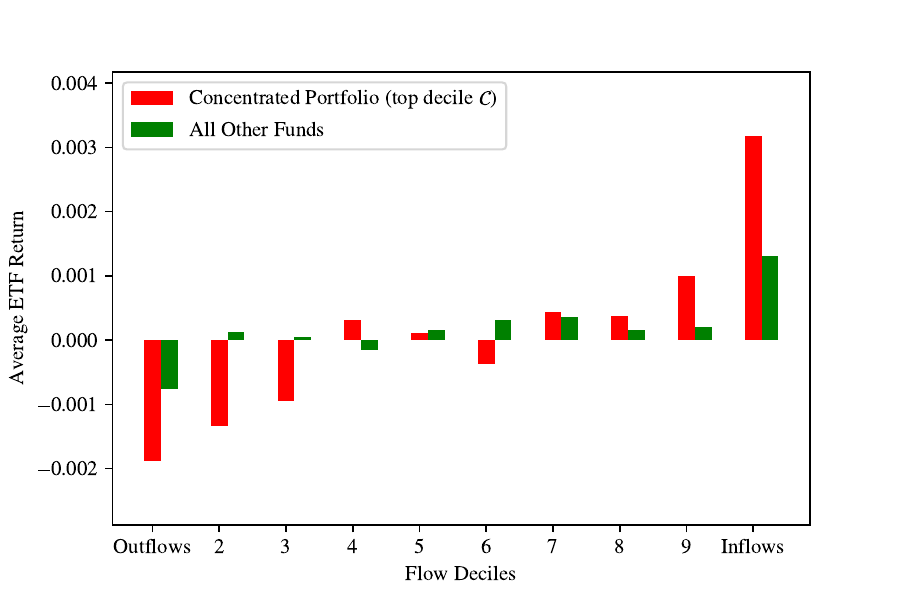}  
    \caption{Alternative Sort: $\frac{F}{V}$ and $\mathcal{C}$}
\end{subfigure}
\label{Portfolio Sorts Robustness}
\end{figure}

\begin{figure}[H]
    \centering
\center
\caption{\textbf{Return Chasing Kernel by ETF types}. The figure plots the cumulative sum of the return chasing coefficients $\{\sum_{s=1}^l\beta_s\}_{l=1}^{L}$ for different estimators and sample splits. Panel a) splits flows by in and outflows. Panel b) splits funds into benchmark trackers and non-benchmarked (active) funds. Panel c) estimates the flow-performance regression via AUM-weighted least squares.  Panel d) provides time-series evidence by replacing the time-fixed effect with a fund-fixed effect. The shaded areas indicate 95\% confidence bands where the cumulative standard errors are clustered at the day and fund level. Cumulative standard errors are computed accounting for the covariance across coefficient estimates, i.e. $\sqrt{\mathbf{1}_t^T \Omega_t \mathbf{1}_t}$ where $\Omega_t$ is the covariance matrix of the coefficients up until lag $t$ and $\mathbf{1}_t$ is a $t \times 1$ vector of ones.}
    \centering
    \begin{subfigure}[t]{0.45\textwidth}
        \centering        
        \caption{Inflows versus Outflows}
       \includegraphics[scale=0.5]{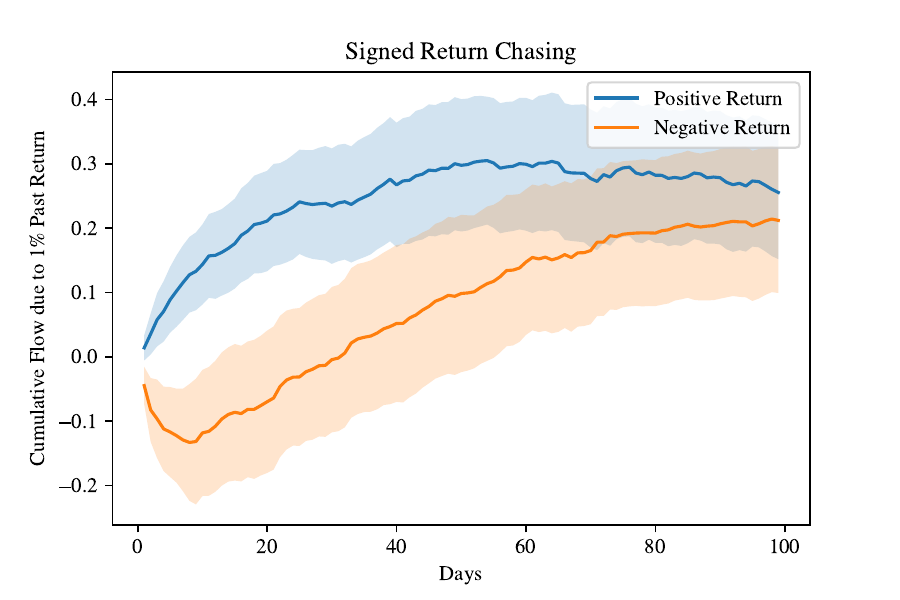}
    \end{subfigure}
    \begin{subfigure}[t]{0.45\textwidth}
        \centering        
        \caption{Active versus Passive}
       \includegraphics[scale=0.5]{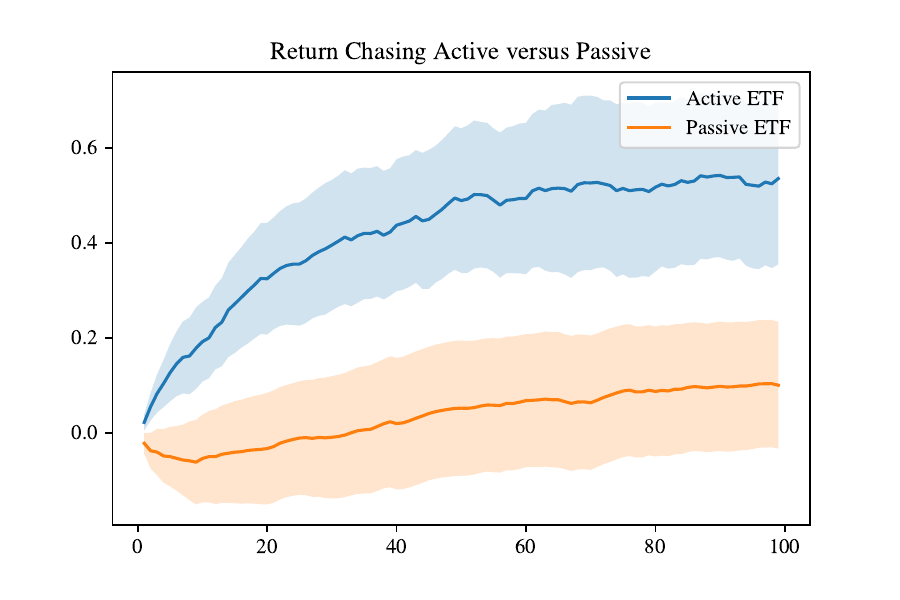}
    \end{subfigure}
    \begin{subfigure}[t]{0.45\textwidth}
        \centering        
        \caption{Aum-Weighted Least Squares}
       \includegraphics[scale=0.5]{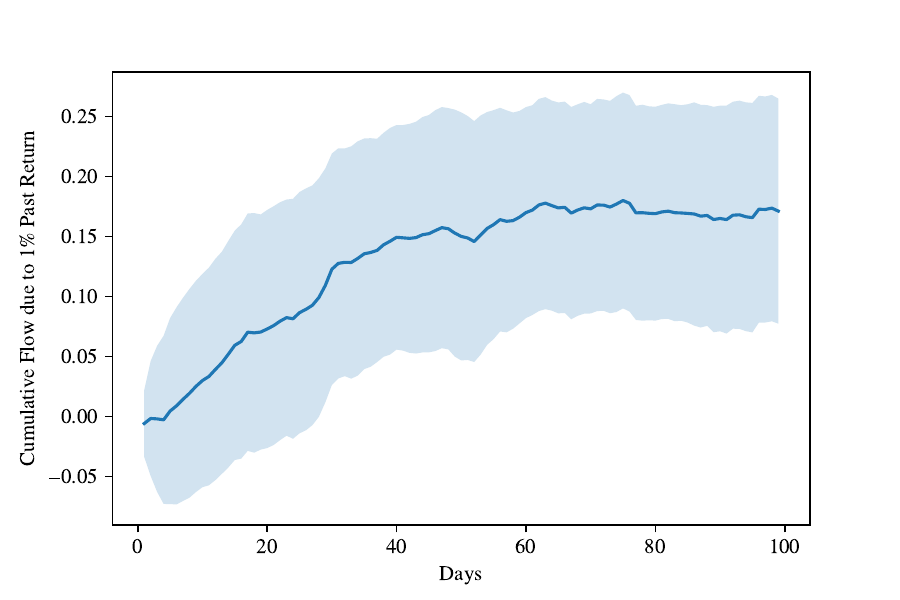}
    \end{subfigure}
    \begin{subfigure}[t]{0.45\textwidth}
        \centering        
        \caption{Fund-fixed Effects}
       \includegraphics[scale=0.5]{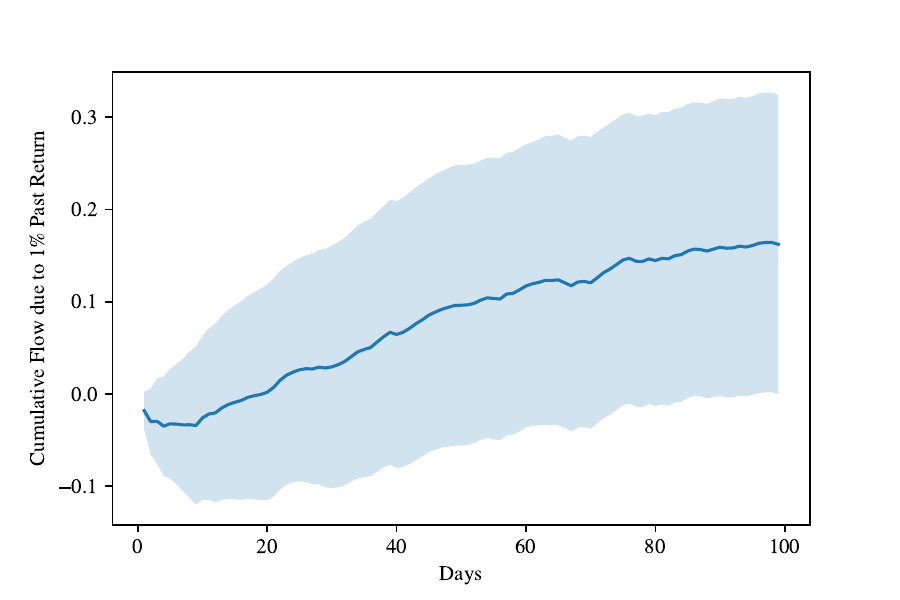}
    \end{subfigure}

    \label{Return Chasing Splits}
\end{figure}

\begin{figure}[H]
\centering
\caption{\textbf{Variance Decomposition: Daily ETF Returns.} The figure plots the explained variance of daily fund returns due to self-incited returns $\beta_i^{ \mathcal{I}} = cov_i(R^{ \mathcal{I}}_{i,t},R_{i,t})/var_i(R_{i,t})$. We group funds by fund size $\mathcal{S}$ and portfolio concentration $\mathcal{C}$ and compute the $\beta_i^{ \mathcal{I}}$. The horizontal axis displays fund size quintiles. The orange bars represent the top 10\% of funds with the most concentrated portfolios within each size quintile. The blue bars represent the remaining 90\%.}
\center
    \includegraphics[scale=0.9]{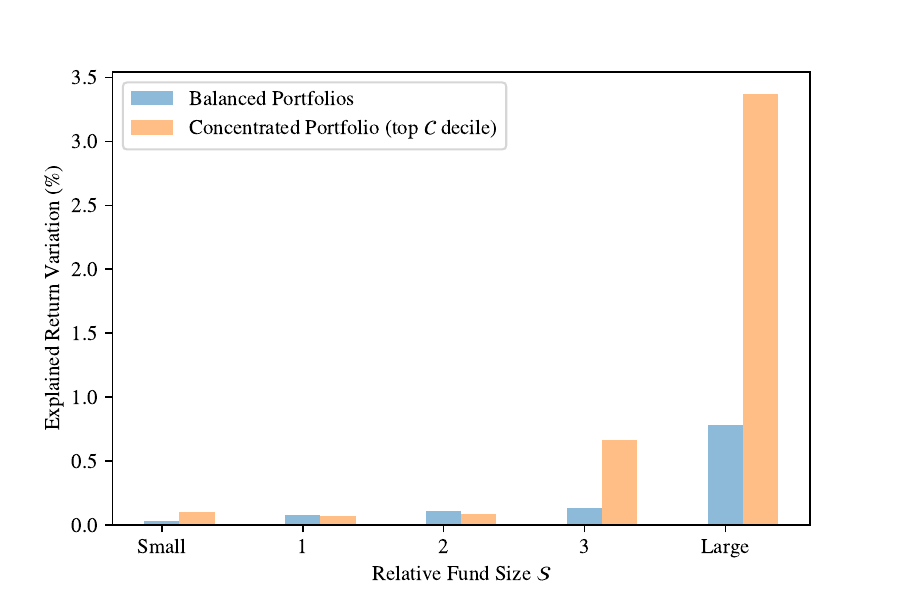}
    \label{Daily Variance Decomposition}
\end{figure}

\begin{table}[H]
\center
\caption{\textbf{Stock-Level Price Impact: Horse Race}. The table reports the estimated price impact coefficients from a horse race of different transformations of stock-level price impact. Specification (1) uses $f_{i,t}^{\eta} \mathcal{I}_{i,n,t-1}$ constructed using $\eta=1$, scaling
demand shocks by market cap, and omitting the volatility pre-factor. Specification (2) uses $f_{i,t}^{\eta} \mathcal{I}_{i,n,t-1}$ constructed using $\eta=1/2$, scaling
demand shocks by market cap, and omitting the volatility pre-factor. Specification (3) uses $f_{i,t}^{\eta} \mathcal{I}_{i,n,t-1}$ constructed using $\eta=1/2$, scaling
demand shocks by daily volume, and including the volatility pre-factor. For simplicity of exposition the table reports $f^{\eta}_{i,t}$ and $\mathcal{I}_{i,n,t-1}$ as $f$ and $ \mathcal{I}$. Standard errors are double-clustered at the stock and fund-day level.}
\begin{tabular}{lllll}
\toprule
{} &      (1)      &      (2)      &      (3)      &      (4)      \\
                      &               &               &               &               \\
\midrule
$f \times  \mathcal{I}$ (Linear)                   &  0.4831***    &               &               &  0.0768       \\
                      &  (3.7541)     &               &               &  (0.5593)     \\
$f \times  \mathcal{I}$ (Linear, Volume)                  &               &  0.1296***    &               &  -0.1552***   \\
                      &               &  (3.7415)     &               &  (-3.7283)    \\
$f \times  \mathcal{I}$ (Sqrt, Volume)             &               &               &  0.3353***    &  0.3987***    \\
                      &               &               &  (7.6032)     &  (6.8188)     \\
\midrule
Effects                 &         Investor-Time         &     Investor-Time    &  Investor-Time  &  Investor-Time   \\
Triple Diff.            &         Yes           &      Yes              &   Yes                 &  Yes         \\

No. Observations      &  34897091     &  34897091     &  34897091     &  34897091     \\
R-squared             &  3.933e-05    &  3.954e-05    &  8.862e-05    &  9.312e-05    \\
\bottomrule
\end{tabular}

\label{HorseRace Stock-Level}
\end{table}

\begin{table}[H]
\center
\caption{\textbf{Fund-Level Price Impact: Horse Race}. The table reports the estimated price impact coefficients from a horse race of different transformations of stock-level price impact. Specification (1) uses $f_{i,t}^{\eta} \mathcal{I}_{i,t-1}$ constructed using $\eta=1$, scaling
demand shocks by market cap, and omitting the volatility pre-factor. Specification (2) uses $f_{i,t}^{\eta} \mathcal{I}_{i,t-1}$ constructed using $\eta=1/2$, scaling
demand shocks by market cap, and omitting the volatility pre-factor. Specification (3) uses $f_{i,t}^{\eta} \mathcal{I}_{i,t-1}$ constructed using $\eta=1/2$, scaling
demand shocks by daily volume, and including the volatility pre-factor. For simplicity of exposition the table reports $f^{\eta}_{i,t}$ and $\mathcal{I}_{i,t-1}$ as $f$ and $\mathcal{I}$. Standard errors are double-clustered at the fund and day level.}
\begin{tabular}{lllll}
\toprule
{} &       (1)      &       (2)      &       (3)      &       (4)       \\
                            &                &                &                &                 \\
\midrule
$f \times  \mathcal{I} (Linear)$    &  1.4791        &                &                &  -2.4266***     \\
                            &  (1.6305)      &                &                &  (-4.5066)      \\
$f \times  \mathcal{I} (Sqrt)$    &                &  0.2397***     &                &  0.1463*        \\
                            &                &  (4.5632)      &                &  (1.7689)       \\
$f \times  \mathcal{I} (Sqrt, Volume)$    &                &                &  0.7762***     &  0.6667***      \\
                            &                &                &  (5.1606)      &  (2.8960)       \\
\midrule
Effects                     &  Entity        &  Entity        &  Entity        &  Entity         \\
                            &  Time          &  Time          &  Time          &  Time           \\
No. Observations            &  1175487       &  1175487       &  1175487       &  1175487        \\
R-squared                   &  0.0004        &  0.0007        &  0.0008        &  0.0008         \\
\bottomrule
\end{tabular}

\label{HorseRace Fund-Level}
\end{table}

\begin{table}[h]
\center
\caption{\textbf{Price Impact Chasing: Active versus Passive}. The table reports the estimated coefficients of pooled OLS regressions of daily flows onto (exponentially weighted) average past returns, as well as decomposed average returns into an impact and a fundamental component.
Specification (1) reports the coefficient to the regression on raw returns $f_{i,t+1} = \alpha_t +  \beta\tilde{R}+ Controls + \epsilon_{i,t+1}$. Specification (2) reports the coefficients to the regression on the decomposed returns $f_{i,t+1} = \alpha_t +  \beta_{1}\tilde{R}^{ \mathcal{I}}_{i,t} + 
     \beta_{2} \tilde{R}^{\perp}_{i,t} + Controls + \epsilon_{i,t+1}$. All specifications control for a constant, lagged flows up to 200 days as well as time-fixed effects. The sample period is 2019 to 2024. Standard errors are double clustered at the day, and the fund level. We split the sample by benchmarked and active (non-benchmark-tracking) funds. All specifications control for a constant, lagged flows up to 200 days, and time-fixed effects. Standard errors are double clustered at the day and the fund level.}
\begin{tabular}{lllll}
\toprule

&\multicolumn{2}{l}{\underline{Benchmarked Funds}} & \multicolumn{2}{l}{\underline{Active Funds}}\\
{} &       (1)       &       (2)       &       (1)       &       (2)       \\
                                &                 &                 &                 &                 \\
\midrule

$R$                   &  0.1746***     &                &  0.8160***      &                 \\
                                &  (2.7785)      &                &  (8.8652)       &                 \\

$R^{ \mathcal{I}}$     &                &  0.1950***     &                 &  0.7734***      \\
                                &                &  (2.9203)      &                 &  (4.2745)       \\
$R^{\perp}$           &                &  0.1804***     &                 &  0.8967***      \\
                                &                &  (2.6787)      &                 &  (8.1457)       \\
\midrule
Controls                        &  Yes           &  Yes           &  Yes           &  Yes           \\

Effects                         &  Time          &  Time          &  Time           &  Time           \\
No. Obs.                &  1050040       &  1050040       &  89525          &  89525          \\
R-squared                       &  0.0140        &  0.0141        &  0.0346         &  0.0351         \\
\bottomrule
\end{tabular}

\label{Impact Chasing AP}
\end{table}

\section{Flow-Induced Trading: ETFs versus Mutual Funds}
\label{Ordinary FIT}
Flow-induced trading by ETF underlies slightly different microstructural mechanics than flow-induced trading by ordinary mutual funds as in \cite{lou2012flow}. We therefore first define the ETF-specific analog of flow-induced trading (FIT), which we label arbitrage-induced trading (AIT).

\subsection{Arbitrage-Induced Trading}
\label{ETFs versus Mutual Funds}
In this section, we outline how ETF flow-driven trading can be constructed from ETF shares outstanding and the stock-level positions held by the ETF. The ETF has $S_{i,t}$ shares outstanding with price $P_{i,t}$, where time (in business days) is denoted by $t$. Total ETF assets under management are $A_{i,t} = P_{i,t}S_{i,t}$. The current trading volume (expressed in shares) of each stock is given by $V_{n,t}$. The return of each stock $n$ is given by $r_{n,t}$. Let $w_{i,n,t}$ denote the portfolio weight of the ETF in stock $n$ and $R_{i,t+1} = \sum_n w_{i,n,t} r_{n,t+1}$ the total return of ETF $i$. We assume that at time $t$ the ETF and the underlying basket have the same price, and therefore the ETF return is given by the weighted sum of the underlying returns. During day $t$, however, the ETF price may deviate from the price of the underlying basket, creating an arbitrage opportunity. For example, high demand for the ETF may drive the ETF price above the underlying. The arbitrage trade implies selling ETF shares (at elevated prices) and buying the underlying. To reduce arbitrage risk, the arbitrageur does not have to bet on the convergence of the spread, but can directly hand the underlying shares to the provider in exchange for ETF shares, which they can sell in the secondary market. 

To this end, the ETF provider specifies to the arbitrageur (authorized participant, AP) the creation basket $\{q^{CU}_{i,n,t}\}^N_{n=1}$ for each creation unit of the ETF (known also as portfolio composition file, PCF). The creation basket constituents $q^{CU}_{i,n,t}$ (denoted in dollars based on current market prices) specify the dollar value of the underlying shares the AP must deliver to the ETF provider in order to receive one ETF share.\footnote{The basket is typically denoted in number of shares. For expositional simplicity (in line with the structural model in the main text) we convert to dollar units using current market prices.} The arbitrage-induced trade of an authorized participant $i$ in stock $n$ is given by $\Delta S_{i,t+1}q^{CU}_{i,n,t}$, where $\Delta S_{i,t+1}$ is the number of ETF shares issued between the close of date $t$ and the close of date $t+1$ and $q^{CU}_{i,n,t}$ the creation basket (denoted in dollars based on current market prices) specified by the ETF provider at the close of date $t$. The creation basket specifies the dollar value of the underlying shares the AP must deliver to the ETF provider in order to receive one ETF share. For example, by delivering $K \times \{q^{CU}_{i,n,t}\}_{n=1}^N$, the authorized participant receives $K$ ETF shares in the primary market and can sell them in the secondary market. Note that the ETF provider has the discretion to supply a trading basket \( q^{CU}_{i,n,t} \) that may differ from the actual constituents \( Q_{i,n,t} \) per unit of ETF share. This flexibility allows the provider to optimize for costs or other operational considerations. However, we have verified that our results remain unaffected by this potential discrepancy. Therefore, for all practical purposes in our analysis, the actual constituents \( Q_{i,n,t} \) can be used, as we do in the main text. To see the equivalence to flow-induced trading, note that the dollar flow $F_{i,t+1}$ into the ETF between $t$ and $t+1$ is the change in ETF shares outstanding multiplied by the price of the ETF $P_{i,t}$, i.e. $F_{i,t+1} = \Delta S_{i,t+1}P_{i,t}$. Similarly, the implied ETF portfolio is given by $w_{i,n,t} = \frac{q^{CU}_{i,n,t}P_{i,n,t}}{\sum_n q^{CU}_{i,n,t}P_{i,n,t}}$. Therefore, the arbitrage-induced trade $\Delta S_{i,t+1}q^{CU}_{i,n,t}$ can also be expressed as $w_{i,n,t}F_{i,t+1}$. Note that $w_{i,n,t}F_{i,t+1}$ is precisely the flow-induced trade by mutual funds, assuming that 100\% of flows are reinvested in line with previous portfolio weights. Aggregating the daily implied arbitrage trades across ETFs and normalizing by supply (market cap $M_{n,t-1}$, yields daily Arbitrage-Induced-Trading (AIT)
\begin{equation}
    AIT_{n,t} = \frac{\sum_i \Delta S_{i,t}q^{CU}_{i,n,t-1}}{M_{n,t-1}}
\end{equation}
which matches the definition of FIT in \cite{lou2012flow}. 

\subsection{AIT's price impact}
We first estimate the price impact of daily arbitrage-induced trades in the following simple contemporaneous panel regression over daily stock returns from 2019 to 2024.
\begin{equation}
    r_{n,t} = \theta_t + \mu_i + \gamma AIT_{n,t} + \epsilon_{n,t}.
\end{equation}
Table \ref{FIT} reports the results.  
\begin{table}[H]
\center
\caption{\textbf{Arbitrage-Induced Trading: Linear Specification}. The table reports the estimated price impact coefficients from pooled OLS regressions using daily data on stock returns and different measures of arbitrage-induced trading $AIT_{n,t}$. Specification (1) to (5) include different combinations of fixed effects. Across all specifications, we control for a constant, momentum and stock-specific volatility. Standard errors are double-clustered at the stock and day level.}
\begin{tabular}{lllll}
\toprule
{} &       (1)      &       (2)      &       (3)      &       (4)      \\
                        &                &                &                &                \\
\midrule
$AIT$                     &  1.5591**      &  1.5555**      &  0.9880**      &  0.9814**      \\
                        &  (2.2604)      &  (2.2579)      &  (2.1381)      &  (2.1248)      \\
\midrule
Controls                &                &  Yes           &  Yes           &  Yes         \\
Effects                 &                &                &  Time          &  Entity        \\
                        &                &                &                &  Time          \\
No. Observations        &  2036923       &  2036923       &  2036923       &  2036923       \\
R-squared               &  0.0014        &  0.0020        &  0.0007        &  0.0008        \\
\bottomrule
\end{tabular}

\label{FIT}
\end{table}
Across all specifications, AIT significantly affects daily contemporaneous returns.
The price impact estimated from arbitrage-induced trading is close to 1 (t-stat 2.1), implying that buying 1\% of the shares outstanding increases stock prices by around 1\%. Because we are estimating price impact at a daily frequency, we also introduce a transformed version of AIT based on equation \ref{Price Impact} and the literature on market impact at a high frequency (\cite{toth2011anomalous}). 
\begin{equation}
    \widehat{AIT}_{n,t} = \sigma_n \sqrt{\frac{\sum_i \Delta S_{i,t}q^{CU}_{i,n,t-1}}{V_{n,t-1}}}
\end{equation}
The transformed version includes scaling arbitrage-induced trades by average daily volume $V_{n,t-1}$, taking a square root, and pre-multiplying by stock-specific volatility $\sigma_n$. If the arbitrage-induced trade is negative, we take the absolute value and sign the transformed  $\widehat{AIT}_{n,t}$.\footnote{Formally $    \widehat{AIT}_{n,t} = \text{sign}(\Delta S_{i,t})\sigma_n \sqrt{\frac{\sum_i |\Delta S_{i,t}|q^{CU}_{i,n,t-1}}{V_{n,t-1}}}$.} Table \ref{FIT Sqrt} reports the results. 
\begin{table}[H]
\center
\caption{\textbf{Arbitrage-Induced Trading: Square Root Specification}. The table reports the estimated price impact coefficients from pooled OLS regressions using daily data on stock returns and different measures of transformed arbitrage-induced trading $\widehat{AIT}_{n,t}$. Specification (1) to (5) include different combinations of fixed effects. Across all specifications, we control for a constant, momentum and stock-specific volatility. Standard errors are double-clustered at the stock and day level.}
\begin{tabular}{lllll}
\toprule
{} &       (1)      &       (2)      &       (3)      &       (4)      \\
                        &                &                &                &                \\
\midrule
$\widehat{AIT}$          &  0.8279***     &  0.8289***     &  0.5502***     &  0.5503***     \\
                        &  (10.451)      &  (10.456)      &  (16.327)      &  (16.261)      \\
\midrule
Controls                 &          &  Yes        &  Yes        &  Yes         \\
Effects                 &                &                &  Time          &  Entity        \\
                        &                &                &                &  Time          \\
No. Observations        &  2036923       &  2036923       &  2036923       &  2036923       \\
R-squared               &  0.0122        &  0.0128        &  0.0061        &  0.0062        \\
\bottomrule
\end{tabular}

\label{FIT Sqrt}
\end{table}
Across all specifications, the price impact estimated from transformed arbitrage-induced trade $\widehat{AIT}_{n,t}$ is 0.55 with considerably higher t-stats of 16. Table \ref{FIT Horserace} runs a horserace between the $\widehat{AIT}_{n,t}$ and $AIT_{n,t}$. We also include $\sqrt{AIT_{n,t}}$ as an additional transformation, to see whether the outperformance of $\widehat{AIT}_{n,t}$ is - beyond the square root transformation - also driven by the alternative scaling (i.e. ADV and the volatility pre-factor). 
\begin{table}[H]
\center
\caption{\textbf{Arbitrage-Induced Trading: Horse Race}. The table reports the estimated price impact coefficients from a horse race of different transformations of arbitrage-induced trading $AIT$. Specifications (1) to (3) compare the estimates in separate univariate specification. Specification (4) considers all transformations jointly. Across all specifications, we control for a constant, momentum and stock-specific volatility. Standard errors are double-clustered at the stock and day level.}
\begin{tabular}{lllll}
\toprule
{} &       (1)      &       (2)      &       (3)      &       (4)       \\
                        &                &                &                &                 \\
\midrule
$AIT$                     &  0.9814**      &                &                &  -0.5009***     \\
                        &  (2.1248)      &                &                &  (-3.2425)      \\
$\sqrt{AIT}$                &                &  0.1571***     &                &  -0.0507**      \\
                        &                &  (16.023)      &                &  (-2.2105)      \\
$\widehat{AIT}$           &                &                &  0.5503***     &  0.7408***      \\
                        &                &                &  (16.261)      &  (8.8815)       \\
\midrule
Controls                 &  Yes        &  Yes        &  Yes        &  Yes         \\

Effects                 &  Entity        &  Entity        &  Entity        &  Entity         \\
                        &  Time          &  Time          &  Time          &  Time           \\
No. Observations        &  2036923       &  2036923       &  2036923       &  2036923        \\
R-squared               &  0.0008        &  0.0047        &  0.0062        &  0.0064         \\
\bottomrule
\end{tabular}

\label{FIT Horserace}
\end{table}
Across all specifications $\widehat{AIT}_{n,t}$ emerges victoriously. In univariate specifications, it achieves the highest $R^2$. In the joint kitchen-sink regression with all transformations, $\widehat{AIT}_{n,t}$ fully subsumes the effect leaving the other impact measures with negative coefficients.

\subsection{Reversal of AIT's price impact}
\begin{figure}[H]
\centering
\caption{\textbf{$AIT$ Impact Reversal}.\\
\footnotesize{The figure plots the cumulative coefficient sum $\beta^{\text{Linear Model}}_s = \sum_{l=0}^{s}\gamma_l$ from the distributed lag model (estimated via OLS)
$r_{n,t} = \theta_t + \mu_i + \sum_{l=0}^{\bar{L}}\gamma_l \widehat{AIT}_{t-l,n} + \epsilon_{n,t}$. We choose 40 days for $\bar{L}$. The model is estimated over the panel of daily stock returns from 2019 to 2024. The shaded region indicates the 95\% confidence interval of the linear model, where standard errors are clustered at the day level.}}
\includegraphics[scale=0.8]{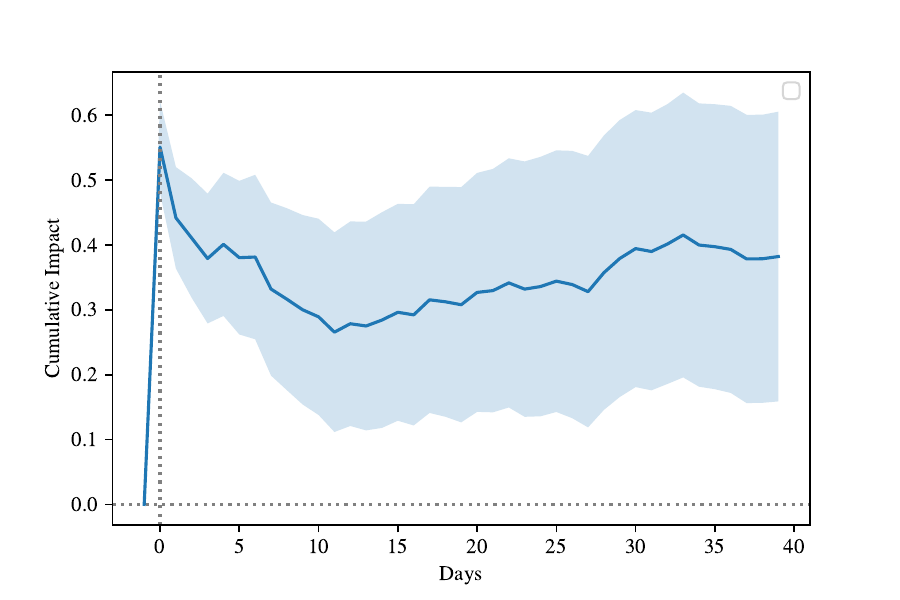}
\label{Impact Decay AIT}
\end{figure}

Does the arbitrage-induced price impact revert over time? Simple regressions of future cumulative returns on lagged AIT are contaminated by the autocorrelation in ETF flows and therefore AIT. A flow today that causes arbitrage-induced price impact is followed by flows tomorrow, therefore masking a potential reversal of the impact. We therefore estimate a distributed lag model over $\bar{L}$=30 lags of AIT 
\begin{equation}
    r_{n,t} = \theta_t + \mu_i + \sum_{l=0}^{\bar{L}}\gamma_l \widehat{AIT}_{t-l,n} + \epsilon_{n,t}.
\end{equation}
Figure \ref{Impact Decay AIT} plots the cumulative sum of the coefficients $\beta_l = \sum_{l=0}^{\bar{L}}\gamma_l$ together with 95\% confidence bands. The cumulative coefficients suggest significant short-term reversal within 10 days following the flow-induced trade. However, after 10 days, the effect stabilizes and around 50\% of the initial impact remains in the price.

\subsection{Factor Structure in ETF demand}
\label{Factor Structure ETF}
In this section, we show that using aggregated flow-induced trading (AIT) does not necessarily deliver an unbiased estimate of price impact $\theta$ if both fundamental returns $r^{\perp}_{t+1,n}$ and ETF flows $f_{i,t+1}$ are driven by a common factor $K_{t+1}$. To this end, note that AIT can be rewritten as $AIT_{n,t} = \sum_i \tilde{Q}_{i,n,t-1} f_{i,t}$ where $\tilde{Q}_{i,n,t-1} = q^{CU}_{i,n,t-1}/M_{n,t-1}$ is ETF ownership relative to shares outstanding and $f_{i,t}=\frac{\Delta S_{i,t}}{S_{i,t-1}}$ is the flow relative to lagged assets. For example, interest cuts may cause a decline in the prices of technology firms. If ETF flows are also affected by this common factor $K_{t+1}$ (here: interest rates) then ETF flows with a technology tilt simultaneously receive outflows. ETF flow-driven buying will therefore be lower for tech stocks, causing a positive correlation between $r_{t+1,n}$ and $AIT_{t+1,n}$ purely because of the latent factor and therefore an upward bias in $\theta$. Formally, assume that $r^{\perp}_{t+1,n} = K_{t+1} \beta_n + \epsilon_{t+1,n}$ and $f_{i,t+1} = K_{t+1} \beta^i + u_{t+1,n}$. Because of the ownership-weighting, the flow exposure to the latent factor $K_{t+1}$ varies across stocks $n$ and is given by $\beta_{n,t}^Q = \sum_i \tilde{Q}_{i,n,t}\beta^i$. If the flow exposure and the return exposure are cross-sectionally correlated $cov(\beta_{n,t}^Q,\beta_n)\neq 0$, then $cov(r^{\perp}_{t+1,n},f^Q_{t+1,n})\neq 0$ and $\theta$ is biased. We can further decompose the bias as
\begin{equation}
    cov(\beta_{n,t}^Q,\beta_n) = \sum_i \beta^i \underbrace{cov_{i,t}(\tilde{Q}_{i,n,t},\beta_n)}_{\text{Factor Tilt}^i_{t}}
\end{equation}
where $\text{Factor Tilt}^i_{t}$ measures the correlation of ETF portfolio tilts and factor exposure, i.e. the extent to which ETF $i$ tilts towards characteristics that drive returns. Therefore, the bias is essentially driven by the extent to which ETF flows are attentive to the portfolio tilts of ETFs. If flows are pure noise, i.e. they are entirely driven by factors unrelated to portfolio holdings, then flow-driven trading uncovers the true $\theta$. If ETF demand - at least to some extent - follows the portfolio tilts of ETFs, then $\theta$ is potentially biased. This concern may be small if ETF flows i) did not obey a strong factor structure and ii) if any commonality is unrelated to portfolio holdings (e.g. only driven by marketing expenses). Both are strongly rejected in the data. 
\begin{figure}[H]
\centering
\caption{\textbf{Factor Structure in ETF Demand}.\\
\footnotesize{The figure plots the cross-sectional variation in weekly ETF flows (share issuances) explained by the first 10 principal components. ETF flows $f_{t+1}^i = S_{i,t+1}/S_{i,t}-1$ are given by the percentage shares in ETF shares outstanding $S_{i,t}$. We winsorize daily and weekly ETF flows at the 99\% level and group them into 10 bins based on their AUM. We then compute value-weighted average flows for each bin and estimate 10 principal components across the 10 value-weighted average flow series.}}
\includegraphics[scale=0.5]{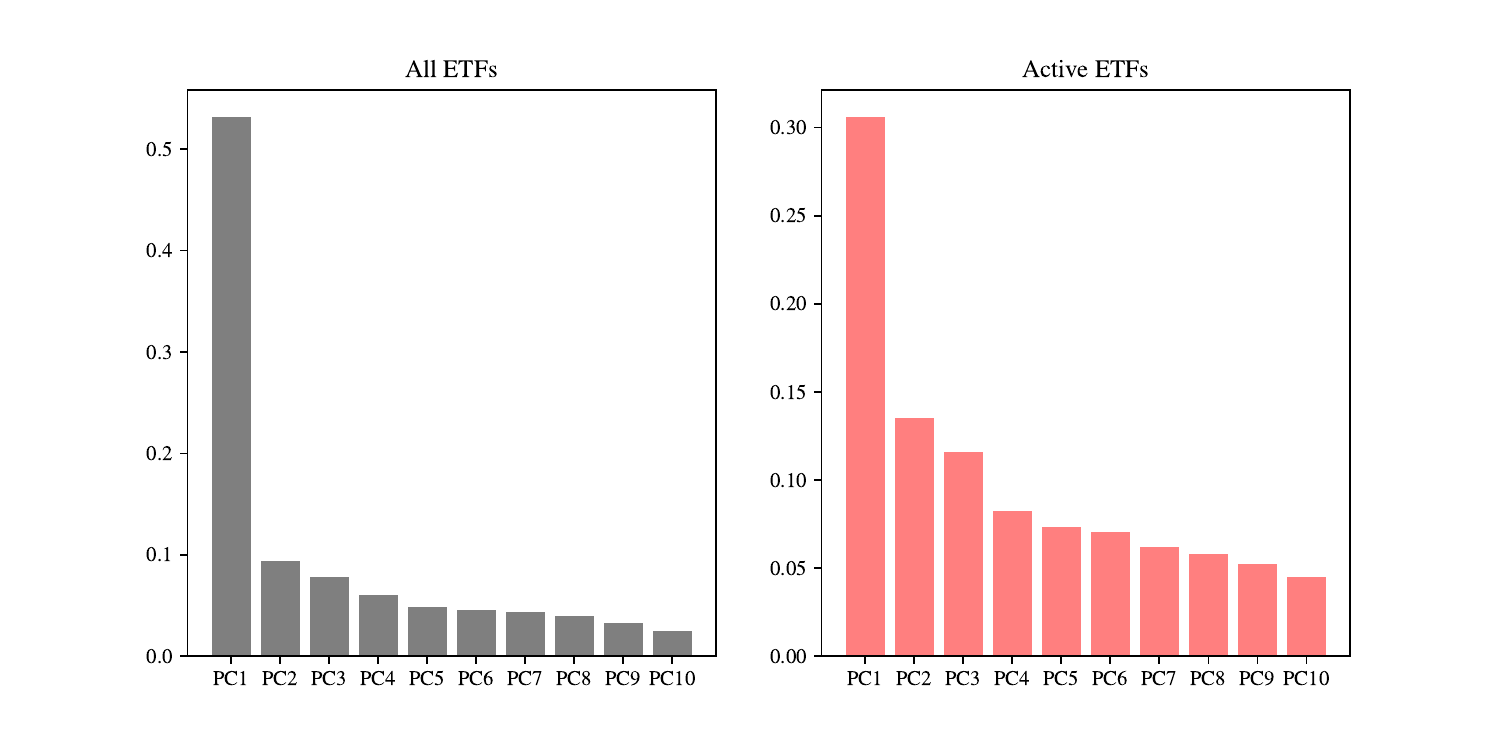}
\end{figure}
We circumvent this by including either fund-time fixed effects (in the stock-level specification) or controlling for raw flows (in the fund-level specification).

\end{appendices}

\end{document}